\documentclass{aa} 

\usepackage{amsmath,amsfonts,amssymb}
\usepackage[dvipsnames]{xcolor}
\usepackage{color}
\usepackage{url}\usepackage[varg]{txfonts}
\usepackage{graphicx}
\usepackage{enumitem} 
\usepackage{array}
\usepackage{dblfloatfix}
\usepackage{multicol}
\usepackage[normalem]{ulem} 
\usepackage{float}  
\usepackage{caption} 

\usepackage{tikz}
\usetikzlibrary{shapes.geometric, arrows}

\usepackage{natbib,twoopt}
\usepackage[breaklinks=true]{hyperref} 
\bibpunct{(}{)}{;}{a}{}{,} 
\makeatletter
\newcommandtwoopt{\citeads}[3][][]{\href{http://adsabs.harvard.edu/abs/#3}%
{\def\hyper@linkstart##1##2{}%
\let\hyper@linkend\@empty\citealp[#1][#2]{#3}}}
\newcommandtwoopt{\citepads}[3][][]{\href{http://adsabs.harvard.edu/abs/#3}%
{\def\hyper@linkstart##1##2{}%
\let\hyper@linkend\@empty\citep[#1][#2]{#3}}}
\newcommandtwoopt{\citetads}[3][][]{\href{http://adsabs.harvard.edu/abs/#3}%
{\def\hyper@linkstart##1##2{}%
\let\hyper@linkend\@empty\citet[#1][#2]{#3}}}
\newcommandtwoopt{\citeyearads}[3][][]%
{\href{http://adsabs.harvard.edu/abs/#3}
{\def\hyper@linkstart##1##2{}%
\let\hyper@linkend\@empty\citeyear[#1][#2]{#3}}}
\makeatother

\usepackage{color}
\hypersetup{colorlinks=true,linkcolor=blue,citecolor=blue,urlcolor=blue}

\usepackage[breaklinks=true]{hyperref}

\makeatletter
\renewcommand*\aa@pageof{, page \thepage{} of \pageref*{LastPage}}
\makeatother

\usepackage[lined,ruled,linesnumbered]{algorithm2e}

\usepackage{algpseudocode}

\newcommand{\bx}{{\bf{x}}}

\newcommand{\bn}{{\bf{n}}}
\newcommand{\bp}{{\bf{p}}}
\newcommand{\bd}{{\bf{d}}}
\newcommand{\bs}{{\bf{s}}}
\newcommand{\bw}{{\bf{w}}}

\newcommand{\bOmega}{\boldsymbol{\Omega}}

\newcommand{\btheta}{\boldsymbol{\theta}}
\newcommand{\bSigma}{{\boldsymbol{\Sigma}}}
\newcommand{\bdelta}{\boldsymbol{\delta}}

\newcommand{\vect}[1]{\boldsymbol{#1}}

\newcommand{\calT}{{\cal{T}}} 
\newcommand{\calM}{{\cal{M}}} 

\usepackage{xcolor}

\definecolor{sgreen}{rgb}{0.19, 0.73, 0.56}
\newcommand{\sop}[1]{{\color{black} #1}}

\tikzstyle{startstop} = [rectangle, rounded corners, minimum width=2cm, minimum height=0.5cm,text centered, text width=7cm, draw=red, fill=red!10]

\tikzstyle{ifloop} = [rectangle, rounded corners, minimum width=2cm, minimum height=0.5cm,text centered, text width=7cm, draw=cyan, fill=white]

\tikzstyle{endstop} = [rectangle, rounded corners, minimum width=\textwidth, minimum height=0.5cm,text centered, text width=\textwidth, draw=red, fill=white]

\tikzstyle{endstop2} = [rectangle, rounded corners, minimum width=0.4*\textwidth, minimum height=0.5cm,text centered, text width=\textwidth, draw=red, fill=white]

\tikzstyle{process} = [rectangle, minimum width=3cm, minimum height=0.5cm, text centered, text width=5.5cm, draw=black, fill=white]

\tikzstyle{process1} = [rectangle, minimum width=1cm, minimum height=0.5cm, text centered, text width=2cm, draw=black, fill=white]

\tikzstyle{process2} = [rectangle, minimum width=2cm, minimum height=0.5cm, text centered, text width=4cm, draw=black, fill=white]

\tikzstyle{process3} = [rectangle, minimum width=2cm, minimum height=0.5cm, text centered, text width=3.5cm, draw=black, fill=white]

\tikzstyle{scheme} = [rectangle, minimum width=3cm, minimum height=0.5cm, text centered, text width=3cm, draw=black, fill=white]

\tikzstyle{losange} = [diamond, minimum width=2cm, minimum height=0.5cm, text centered, draw=black, fill=white]

\tikzstyle{textt} = [rectangle, minimum width=5cm, minimum height=0.5cm, text centered, text width=6cm, draw=white, fill=white]

\tikzstyle{textt2} = [rectangle, minimum width=5cm, minimum height=0.5cm, text centered, text width=0.5cm]

\tikzstyle{arrow} = [thick,->,>=stealth]
\tikzstyle{line} = [thick,-,>=stealth]

\tikzstyle{MCloop} = [thick,-,>=stealth]

\tikzstyle{header} = [rectangle, minimum width=\textwidth, minimum height=0.5cm, text centered, text width=\textwidth, draw=black, fill=white]

\begin{document} 

\title{Semi-supervised standardized detection of extrasolar planets }

 \author{S. Sulis \inst{1}
 \and D. Mary \inst{2}
 \and L. Bigot \inst{2} 
 \and M. Deleuil \inst{1} 
 }
 \institute{    
 Universit\'e Aix Marseille, CNRS, CNES, LAM, Marseille, France\\
\email{sophia.sulis@lam.fr}
   \and
 Universit\'e C\^ote d'Azur, Observatoire de la C\^ote d'Azur, 
                CNRS, Lagrange UMR 7293, CS 34229, 
 06304, Nice Cedex 4, France
 }


 \abstract
 {The detection of small exoplanets with the radial velocity (RV) technique is limited by various poorly known noise sources of instrumental and stellar origin. {As a consequence, current detection techniques often fail to provide reliable estimates of the significance levels of detection tests in terms of false-alarm rates or $p$-values.}
 }
  {We designed an RV detection procedure that provides reliable $p$-value estimates while accounting for the various  noise sources typically affecting RV data. The method is able to incorporate ancillary information about the noise (e.g., stellar activity indicators) and specific data- or context-driven data (e.g. instrumental measurements, magnetohydrodynamical simulations of stellar convection, and simulations of meridional flows {or magnetic flux emergence}).
 }
 {The detection part of the procedure uses a detection test that is applied to a standardized periodogram. Standardization allows an autocalibration of the  noise sources with partially unknown statistics (algorithm 1). The estimation of the $p$-value of the test output is based on dedicated Monte Carlo simulations that allow handling unknown parameters (algorithm 2).  The procedure is versatile in the sense that the specific pair (periodogram \sop{and} test) is chosen by the user. Ancillary or context-driven data can be used if available.
} 
{We demonstrate by extensive numerical experiments on synthetic and real RV data from the Sun and $\alpha$CenB that the proposed {method reliably allows estimating the $p$-values.} The method also provides a way to evaluate the dependence of the estimated {$p$-values} that are attributed to a reported  detection on modeling errors. It is a critical point for RV planet detection at low signal-to-noise ratio to evaluate this dependence. {The python algorithms developed in this work are available on GitHub.} 
 }
{{Accurate estimation of $p$-values when unknown parameters are involved in the detection process is an important but only recently addressed question in the field of RV detection.  Although this work presents a method to do this,  the statistical literature discussed in this paper may trigger the development of other strategies.}} 
\keywords{< Techniques: radial velocities - stars: activity - Planets and satellites: detection - Methods: statistical>}

 \maketitle

\section{Introduction }
\label{sec1}
 

When radial velocity (RV) time series are analysed, exoplanet detection tests aim at deciding whether the time series contains one or several planetary signatures plus noise (the alternative hypothesis,  noted ${\mathcal{H}}_1$), or  only noise  (the  null hypothesis, noted ${\mathcal{H}}_0$). With the recent advent of  very stable and high-precision spectrographs (\citeads{2010SPIE.7735E..0FP}; \citeads{2016SPIE.9908E..6TJ}), instrumental noise has reached levels that are so low \citepads{2016PASP..128f6001F} that the noise generated by stellar activity has become  the limiting factor for the detection of Earth-like planets. Stellar noise is frequency dependent (colored) because stellar variability results from several phenomena that contribute simultaneously to the observations and evolve at different timescales. The main sources of stellar noise that have been identified so far include magnetic cycles, starspots, faculae, inhibition of the convective blueshift by plages, meridional circulation, convective motions of the stellar atmosphere (granulation and supergranulation), and acoustic oscillations.  The overall stellar noise affects the frequency range in which planets can be found and may  dominate the $\sim10$ cm/s amplitude level that is typical of an Earth-like exoplanet orbiting a Sun-like star at $1$ AU.

The development of techniques to mitigate stellar activity for the study of exoplanets is an active field of research. Several methods and tools have been developed to account for these different sources in the RV time series (see \citetads{2021arXiv210406072M} for a recent review). 
First, some specific observation strategies  are commonly used to reduce the contribution of short-timescale activity components (\citeads{2011ApJ...743...75H}; \citeads{2011A&A...525A.140D}).
Second, to correct for variability that evolves at the stellar rotation period (spots and plages), the most basic technique consists of fitting a model to the RV dataset (e.g., sinusoidal functions at the stellar rotation period \citepads{2011A&A...528A...4B}, or {quasi-periodic} Gaussian processes ({see, e.g., \citeads{2012MNRAS.419.3147A}; \citeads{2011CeMDA.111..235B}; \citeads{2013MNRAS.429.2052B};} \citeads{2015MNRAS.452.2269R}; \citeads{10.1214/21-AOAS1471}; \citeads{2021MNRAS.507.1847R})).
To constrain the parameters of these models, activity indicators (e.g.,  the S-index \citepads{1968ApJ...153..221W}, ${\rm log R'_{HK}}$ \citepads{1984ApJ...279..763N}, or the bisector span \citepads{2001A&A...379..279Q}) are commonly used if correlations with the RV dataset are found. 
Some relations with photometry (\citeads{2012MNRAS.419.3147A}; \citeads{2018MNRAS.480L..48Y}) 
or the study of the chromatic dependence of the stellar activity (\citeads{2017A&A...607A...6M}; \citeads{2018A&A...614A.122T}; \citeads{2018A&A...620A..47D}) can also be used.
Recent advances in RV extraction processes from stellar spectra in particular are very promising (\citeads{2017ApJ...846...59D}; \citeads{2021MNRAS.505.1699C}; \citeads{2021A&A...653A..43C};  \citeads{2020arXiv201100003D}).

However, despite the diversity of existing techniques to reduce the stellar noise, residual noise sources at unknown levels translate into uncertainties in the interpretation of the RV detection tests. 
 In practice, it is difficult to provide a reliable estimation of the significance level of a reported detection. For this reason, examples of a debated detection are numerous (see as examples the cases of HD 41248 b and c (\citeads{2014ApJ...794..110J}; \citeads{2014A&A...566A..35S}),  HD 73256 b (\citeads{2003A&A...407..679U}; \citeads{2018AJ....156..213M}),  AD Leo b (\citeads{2018AJ....155..192T}; \citeads{2020A&A...638A...5C}), Aledebaran c (\citeads{2015A&A...580A..31H}; \citeads{2019A&A...625A..22R}), or Kapteyn b and c (\citeads{2014MNRAS.443L..89A}; \citeads{2021AJ....161..230B})). 
We discuss the particular case of $\alpha$CenBb at the end of this paper as an example  (\citeads{2012Natur.491..207D}; \citeads{2013ApJ...770..133H}; \citeads{2016MNRAS.456L...6R}; \citeads{2021arXiv210514222T}).

The main question we address here is the evaluation of reliable significance levels of detection tests in the presence of unknown colored noise. Significance levels are often interpreted through analyses of false-alarm probabilities (FAP) or of $p$-values. The former are fixed before the tests are conducted, while the latter measure a degree of surprise of the observed test statistic under the null hypothesis. In this work, we consider $p$-values. The estimation of these values has been amply discussed in the literature for the case when the noise model contains several unknown parameters (\citeads{10.2307/2669749}). {Connections of our approach with the literature are discussed in Sec. \ref{connec}.} An accurate estimation of the $p$-values is critical for designing reliable detection methods of small-planet RV signatures. \\


Because the sampling grid is irregular, an RV detection is most often performed through the Lomb-Scargle periodogram (LSP; \citeads{1982ApJ...263..835S}) or variants of the periodograms  (see, e.g.,  \citeads{1981AJ.....86..619F}, \citeads{1999ApJ...526..890C}, \citeads{2004MNRAS.354.1165C}, \citeads{2007A&A...467.1353R}, \citeads{2007A&A...462..379B}, \citeads{2009A&A...496..577Z}, \citeads{2013A&C.....2...18B}, \citeads{2014MNRAS.441.2253J}, \citeads{2014MNRAS.441.1545T},   \citeads{2016MNRAS.458.2604G},  \citeads{2017MNRAS.464.1220H}, \citeads{2022A&A...658A.177H}).
While the marginal distribution of the periodogram components at each frequency is often known for white Gaussian noise (WGN) of known variance (e.g., a $\chi^2$ distribution for the LSP), the joint distribution of the periodogram components is much more difficult to characterize because the components exhibit dependences that are dictated by the sampling grid (or equivalently, by the spectral window). 

In practice, specific noise models are applied to correct RV data for the activity, and the $p$-values of the test statistics are evaluated on the RV residuals assuming they follow the statistics of a WGN, or that they have a known covariance matrix. 
When the noise is considered WGN, some methods are based on approximate analytical expressions \citepads{2008MNRAS.385.1279B}. Other methods compute the FAP using the formula that would be obtained for independent components, where  an ad hoc number of independent frequencies is plugged in (see \citeads{1986ApJ...302..757H}, \citeads{1998MNRAS.301..831S}, \citeads{2004MNRAS.354.1165C}, as well as \citeads{2014MNRAS.440.2099S} and \citeads{2015MNRAS.450.2052S} for a critical analysis of this approach). Finally, another family of approaches is based on Monte Carlo or bootstrap techniques (\citeads{2004A&A...420..789P}, \citeads{2012IAUS..285...81S}), sometimes aided by results from extreme-value distributions (\citeads{2014MNRAS.440.2099S}, \citeads{2015MNRAS.450.2052S}). 

In the large majority of cases, however, the noise in RV residuals is not white, but colored \sop{and with} an unknown power spectral density (PSD). Providing reliable FAP estimates in this situation remains, to our knowledge, an open problem. The assumption that the noise PSD is flat leads logically to increased false alarms in the PSD bumps, and to a loss of detection power in the PSD valleys (e.g., see Fig. 2 in \citeads{2016arXiv160107375S}).
 This problem is only partially solved in practical approaches where a covariance model is fitted to the data and used to produce residuals that are assumed to be white. For instance, \citetads{2020A&A...635A..83D} recently presented an analytical formulation of the $p$-value  resulting from the generalization of \citetads{2008MNRAS.385.1279B} to the case of colored noise when the covariance matrix is known. Their results indicate that the FAP estimates are highly dependent on the considered covariance matrix, with variations directly related to the mismatch between the true and assumed covariance matrix. Similar effects were analyzed in \citetads{2016arXiv160107375S} and in Sec. 3.5 of \citet{sulis:tel-01687077}.
{To our knowledge, all methods aimed at evaluating the FAP or $p$-values associated with periodogram tests assume that the noise is eventually white, either directly or because the model used for the covariance is considered exact, so that the signal can be whitened}. Since estimation errors in the noise statistics lead to fluctuations in the  {$p$-value estimates}, our incomplete knowledge of the noise sources and their statistics represents a fundamental limitation to how much we can trust these estimates. The present work deals precisely with this problem.
Another critical point is the consistency of the {$p$-value} estimate when different noise models are used in the detection process (this point was very clearly raised by  \citeads{2016MNRAS.456L...6R}, for instance). The controversy about a low-mass planet detection often resides in the different noise models that are assumed for the data. Because i) noise models can never be exact and ii) the estimated {$p$-values} are model dependent, incorporating model errors in the estimation of these values appears to be a relevant (but new) approach, and we follow this path below.  \\


In this paper, we propose a new RV planet detection procedure that can generate reliable estimates of the significance levels of detection tests by means of $p$-values. The approach is built on a previous study \citepads{2020A&A...635A.146S}, which was limited to the case of regularly sampled time series and primarily considered the effect of granulation noise. Both of these limitations are removed in the present paper. \\
The approach is flexible and can deal with the main characteristics that are encountered in practice with RV detection:  irregular sampling and various noise sources with partially unknown noise statistics. 
The basic principle of the procedure is to propagate errors on the noise model parameters to derive accurate $p$-values of detection tests by means of intensive Monte Carlo simulations.
The choice of the noise models involved in the procedure is entirely left to the user, as is the choice of the periodogram and detection test. 
We note that without loss of generality, the procedure can be applied to an RV dataset resulting from any extraction process (including, e.g., the extraction techniques recently described in  \citeads{2021MNRAS.505.1699C} or  \citeads{2021A&A...653A..43C}).
In addition, the procedure can exploit training time series of stellar or instrumental noise, if these  data are available, for instance, through auxiliary observations or realistic simulations. \\
The core of the new detection approach is presented in Sec~\ref{sec2} (algorithm~\ref{fig_algo1}). %
The numerical procedure for estimating  the $p$-values of the test statistics produced by algorithm~\ref{fig_algo1} is presented in Sec~\ref{sec3} (algorithm~\ref{fig_algo3}). 
Validations of the $p$-value estimation procedure under different configurations are given in Sec. \ref{sec4} and \ref{sec5}.
We conclude with the application of the detection procedure to the RV data of $\alpha$CenB, where an Earth-mass planet detection has been debated (Sec.~\ref{sec6}). In this last section, we show that the proposed detection procedure can be used to evaluate the robustness of a planet detection with respect to specific model errors.
Examples of how to use the algorithms we developed in this work are available on GitHub\footnote{\url{https://github.com/ssulis/3SD}}, and practical details are given in Appendix.~\ref{appA}.

\section{Semi-supervised standardized detection}
\label{sec2}

Our notations are as follows: Bold small (capital) letters without subscript indicate column vectors (matrices). Vector (matrix) components are denoted by subscripts. For instance, we write vector $\bx$ as  $\bx=[\bx_1,\hdots \bx_N]^\top$. We denote by $\widehat{z}$     the estimate of a quantity $z$. Throughout the paper, the notation | means  ``conditional on'' (and it is not necessarily used for probability densities).
Appendix~\ref{ann1} summarizes the main notations of this work.

\subsection{{Hypothesis testing}}
\label{sec21}

We consider an RV time series under test, $\bx$,  irregularly sampled at $N$ time instants ${\bf{t}}=[t_1,t_2,\hdots,t_N]^\top$. 
Our detection problem can be cast as a binary hypothesis problem of the form

\begin{equation} 
 \left\{ 
 \begin{aligned}
        {\cal H}_0~: \bx &= \bd|{\cal{M}}_\bd(\boldsymbol{\theta}_\bd) + \bn, \\
        {\cal H}_1~: \bx &= {\bf{s}}|{\cal{M}}_\bs(\boldsymbol{\theta}_\bs)+  {\bf{d}}|{\cal{M}}_\bd(\boldsymbol{\theta}_\bd) + {\bf{n}}. \\
 \end{aligned}
 \right.
 \label{hyp}
\end{equation}

The  signal $\bd$ accounts for various effects that typically affect RV data. A nonexhaustive list of such effects includes long-term instrumental trends, stellar magnetic activity, or {RV trends due to an additional companion in the system. This nuisance} signal, which is the unknown mean under ${\cal H}_0$, cannot be estimated by numerical simulations.  While $\bd$ can be of random origin, it is customary to model it as a deterministic signature (see, e.g.,  \citeads{2012Natur.491..207D}). This is our approach in the following. We refer to $\bd$ as a nuisance signal and write generically that $\bd $ is generated from some model ${\calM}_\bd$ with  parameters $\btheta_\bd$.

The noise $\bn\sim{\cal{N}}({\bf0},\boldsymbol{\Sigma})$ is a zero-mean stochastic noise of unknown covariance matrix $\boldsymbol{\Sigma}$ that we decompose into 
\begin{equation} 
\boldsymbol{\Sigma} = \boldsymbol{\Sigma}_c+{\boldsymbol{\Sigma}_w}, 
 \label{eq_n}
\end{equation}
with $ \boldsymbol{\Sigma}_w$ a diagonal covariance matrix whose entries account for possibly different uncertainties on the RV measurement.  If they are equal, $ \boldsymbol{\Sigma}_w = \sigma_{\bw}^2\boldsymbol{I}$ , with $ \sigma_{\bw}^2$ the variance  of a WGN (noted $\bw$ below), and  $\boldsymbol{\Sigma}_c$ the unknown covariance matrix of a colored noise \sop{We note that, in this paper,} $\Sigma_c$ is considered fixed. Temporal variations of  $\Sigma_c$ such as those considered by \citetads{2015MNRAS.446.1493B} might in principle  be considered in our numerical procedure by sampling according to a time-dependent covariance structure, but this is beyond the scope of this paper. A nonexhaustive list of effects that can be included in $\bn$ entails meridional circulation, magnetic flux emergence, stellar granulation, supergranulation, and instrumental noise. The choice of which type of RV noise component fits in $\bd$ or $\bn$ depends on the assumptions made by the user about the RV time series under test.

Under the alternative hypothesis (one or more planets are present), time series $\bx$ contains 
the same noise components $\bd$ and $\bn,$ plus a deterministic signal ${\bf{s}}$ corresponding to the Keplerian signatures of planets (model ${\cal{M}}_\bs$). The number of planets and their Keplerian parameters $\boldsymbol{\theta}_\bs$ are unknown. \\


Throughout this paper, we consider the optional case in which context-driven data of the stochastic colored noise $\bn$ in Eq. \eqref{hyp} can be made available through realistic simulations, for instance. We call these data the null training sample (NTS). By construction, this NTS is free of any possible contamination by the planetary signal.

For example, focusing on the stellar granulation noise, \citetads{2020A&A...635A.146S} showed that 3D magnetohydrodynamical (MHD) simulations can generate realistic synthetic time series of this RV noise source (see also Appendix~\ref{appB}). This was also demonstrated in \citetads{2022AJ....163...11P} with synthetic stellar spectra simulators. This suggests that using simulations of stellar granulation in the detection process could provide an alternative to the observational strategy of binning spectra over a night \citepads{2011A&A...525A.140D}. \citetads{2015A&A...583A.118M} and \citet{2019A&A...625L...6M} showed that while convection (granulation and supergranulation) dominates at high frequencies, where it acts as a power-law noise ($\sim 1/f$), convection has energy at all frequencies, so that averaging on short timescales is not efficient enough to detect RV signatures of Earth-like planets that are hidden in the convection noise. 

Other examples in which NTS can be obtained are 3D MHD simulations of stellar supergranulation \citep[e.g.,][]{2018LRSP...15....6R}, empirical simulations of stellar meridional flows \citep{2020A&A...638A..54M}, or specific data-driven observations representative of the instrumental noise (e.g., microtelluric lines; \citeads{2014A&A...568A..35C} \citeads{2010A&A...524A..11S}) affecting the RV data. In general, all realistic synthetic time series of the stochastic colored noise sources affecting the RV data  can be included in the NTS.

{ This setting poses the general question how this NTS might be leveraged to improve the detection process, both in terms of detection power and of the accuracy of the computed $p$-value. 
This approach was proposed and analyzed in \citetads{2017ITSP...65.2136S}, but with ${\bd}=\bf{0}$ and a regular sampling for $\bx$ in model \eqref{hyp}.}  In this approach, the  detection test was automatically calibrated with the NTS through a  standardization of the  periodogram.   
While this preliminary work has demonstrated improved performances in terms of power and control of the $p$-value, it presents two  important limitations for RV detection purposes. First, the approach accounts only for noise that can be generated through realistic simulations. Second, the scope was limited to regular sampling. 
In contrast, the  detection procedure proposed below can cope with all components  of  noise, including  instrumental and  stellar  noises of different origins {(${\bd}\neq \bf{0}$)} and  any type of time sampling. 
This general procedure is semi-supervised in the sense that it incorporates ancillary sources of noise that affect the data. Because it is also based on a periodogram standardization, we call this procedure semi-supervised standardized detection (\textsf{3SD} for short). {The \textsf{3SD} procedure contains two algorithms: one for the detection, and one for evaluating the resulting significance levels}.

\subsection{{Periodogram standardization}}
 
\subsubsection{{An NTS is available}}
We assume that the NTS,  if available,  takes the form of  $L$ sequences of synthetic time series of noise $\bn$ in Eq. \eqref{hyp},
\begin{equation}
{\cal{T}}_L:=\{\bn^{(i)} \}, i=1,\hdots, L,\quad \bn^{(i)} \sim{\cal{N}}({\bf{0}},\bSigma).
\label{eq_nts}
\end{equation}
An efficient way to use this side information in the detection process is to compute standardized periodograms (see \citeads{2020A&A...635A.146S}). 
Let $\Omega$ denote the set of frequencies at which the periodogram, noted $\bp$, is computed ($\Omega$ and the type of periodogram $\textsf{P}$ are defined by the user). We note $\nu_k$ these frequencies and  $N_\Omega$ their number.  The periodograms of each of the $L$ series of the NTS, $\bp_{\ell}(\nu_k)$,  with $\ell=1,\cdots, L$ can be used to
 form an averaged periodogram, $\overline{\bp}_L(\nu_k)$, as the sample mean of these $L$ periodograms,
\begin{equation} 
        \overline{\bp}_L(\nu_k )  : =  \frac{1}{L} \sum_{\ell=1}^{L}  \bp_{\ell}(\nu_k).
        \label{avper} 
\end{equation}
 The standardized periodogram is then defined as
\begin{equation} 
        \widetilde{\bp}(\nu_k):= \frac{\bp(\nu_k)}{\overline{\bp}_L(\nu_k) }.
        \label{eq_Ptilde} 
\end{equation}

In the following,  $\bf p$ and  ${\bf{\overline{p}}}_L$  denote the vectors concatenating the periodogram ordinates at all frequencies in $\Omega,$ and the standardized periodogram is noted ${{\bf{\widetilde{p}}\;|\;{\bf{\overline{p}}}}}_L$ or simply ${\bf{\widetilde{p}}}$.

In the case of  regular sampling  and when $\bd=\bf{0}$, \citetads{2020A&A...635A.146S} demonstrated that standardization avoids both a loss of detection power in valleys of the noise PSD and an increase in false alarm rate in its peaks (see \citetads{2016arXiv160107375S}, and Sec.3.5.3 of \cite{sulis:tel-01687077}).
{An important point of the present paper is to show that these desirable features also hold in the case of irregular sampling for different models of $\bd$ and $\bn$ (see Sec.~\ref{sec4} and Sec.~\ref{sec5}).} 

\subsubsection{No NTS is available}

Simulations (e.g., 3D MHD simulations of stellar granulation) are computationally demanding, and generating very long time series (months or years) may therefore appear beyond reach for standard RV planet searches. {It is also worth mentioning that helioseismology has revealed that 3D MHD simulations are unable to reproduce the power of large-scale flows on the Sun (supergranulation), which means that improvements are still needed in this field of research \citepads{2012PNAS..10911928H}}.
For these reasons, we also consider the case in which no NTS of $\bn$ in \eqref{hyp} is available. In this case, the periodogram standardization relies on a parametric model ${\cal{M}}_\bn$ whose parameters are estimated from the RV data. \sop{We note that,} when an NTS is available, model ${\cal{M}}_\bn$ is not necessary for the detection process in algorithm \ref{fig_algo1}. Therefore, it was not made explicit in \sop{the notation used} in model \eqref{hyp}. In this approach, synthetic time series of $\bn$ are  generated and used to compute the averaged periodogram used in Eq. \eqref{avper}.
Numerous examples of stochastic parametric models exist, and a (nonexhaustive) relevant list for RV data includes Harvey-like functions \citepads{1985ESASP.235..199H}, autoregressive (AR) models (\citeads{Brockwell_1991}; \citeads{2019A&A...627A.120E}), or Gaussian processes \citepads{Rasmussen:2005:GPM:1162254}). 

A particular noise configuration happens when noise $\bn$ is white ($\boldsymbol{\Sigma}_c=0,$ leading to $\boldsymbol{\Sigma} = \sigma^2\boldsymbol{I}$ in \eqref{eq_n}). In this case, the standardization of the periodogram in Eq. \eqref{eq_Ptilde} is simply performed by the estimated variance of the RV time series.

\subsection{Considered detection procedure}
\label{sec24}

\begin{figure}[t!]
\begin{center}
\hspace{-2cm}
\begin{tikzpicture}[node distance=1.5cm] 



\node (b0) [textt, yshift = -1.2cm,xshift=0cm] {};

\node (start) [startstop,below of=b0,align=justify, yshift = 0cm,text width=6.2cm] 
{
INPUTS \\ 
\vspace{-0.2cm}\\
${\bf x}$: time series under test\\
Selected couple (periodogram $\textsf{P}$, test $\textsf{T}$) \\
$\Omega$: considered set of frequencies\\
${\cal{T}}_L$: null training sample of ${\bf{n}}$ (if any)\\
${\cal{M}}_\bn$: parametric model for ${\bf{n}}$ (if any)\\
${\cal{M}}_\bd$: parametric model for ${\bf{d}}$ (if any)\\
${\bf c}$: activity indicators time series (if any) 
}; 

\draw [cyan,fill=cyan!15] (-3.3,-8.2) -- (3.3,-8.2) -- (3.3,-5.2) -- (-3.3,-5.2)-- cycle; 
\draw [magenta,fill=magenta!10] (-3.3,-9.7) -- (3.3,-9.7) -- (3.3,-15.9) -- (-3.3,-15.9)-- cycle; 
\draw [magenta,fill=magenta!50] (-3.2,-10.5) -- (3.2,-10.5) -- (3.2,-12.2) -- (-3.2,-12.2)-- cycle; 
\draw [magenta,fill=magenta!50] (-3.2,-12.8) -- (3.2,-12.8) -- (3.2,-15.7) -- (-3.2,-15.7)-- cycle; 
\draw [magenta,fill=magenta!10] (-3.3,-16.4) -- (3.3,-16.4) -- (3.3,-19.1) -- (-3.3,-19.1)-- cycle; 

\node (l1) [ifloop, below of=start, yshift = -1cm, text width = 6.5cm,minimum width=4.5cm,fill=cyan!15] 
{\textbf{If} there is a nuisance signal ${\bf{d}}$ (${\cal{M}}_\bd \neq \emptyset$)};
\draw [arrow] (start) -- (l1); 
\node (t2) [textt2, left of=l1, xshift=-2.1cm, text width=1cm] 
{1.};
 
\node (l2) [process, below of=l1, yshift = 0.4cm, text width = 6.2cm,minimum width=4.5cm, draw=cyan] 
{Estimate $\widehat{\boldsymbol{\theta}}_\bd$, the parameters  of model ${\cal{M}}_\bd$ \\  based on the input data ${\bf{x}}$ (and ${\bf c}$ if any)};
\draw [arrow] (l1) -- (l2); 
\node (t2) [textt2, left of=l2, xshift=-2.1cm, text width=1cm] 
{2.};

\node (l3) [process, below of=l2, yshift = 0.2cm, text width = 6.2cm,minimum width=4.5cm, draw=cyan] 
{Compute the data residuals as\\
${\bf{x}} \leftarrow {\bf{x}}-\widehat{{\bf{d}}}|{\cal{M}}_\bd(\widehat{\boldsymbol{\theta}}_\bd)$};
\draw [arrow] (l2) -- (l3); 
\node (t2) [textt2, left of=l3, xshift=-2.1cm, text width=1cm] 
{3.};

\node (l4) [process, below of=l3, yshift = 0.3cm, text width = 6cm,minimum width=4.5cm] 
{Compute periodogram $\bf p({\bf{x}})$ with \textsf{P}};
\draw [arrow] (l3) -- (l4); 
\node (t2) [textt2, left of=l4, xshift=-2.1cm, text width=1cm] 
{4.};

\node (l5) [ifloop, below of=l4, yshift = 0.6cm, text width = 6.5cm,minimum width=4.5cm, draw=magenta,fill=magenta!10] 
{\textbf{If} the  noise ${\bf{n}}$ is colored};
\draw [arrow] (l4) -- (l5); 
\node (t2) [textt2, left of=l5, xshift=-2.1cm, text width=1cm] 
{5.};

\node (l6) [ifloop, below of=l5, yshift = 0.7cm, text width = 5.3cm,minimum width=4.5cm, draw=magenta,fill=magenta!50] 
{\textbf{If} NTS available (${\cal{T}}_L \neq \emptyset$)\\};
\node (t2) [textt2, left of=l6, xshift=-2.1cm, text width=1cm] 
{6.};

\node (l7) [process, below of=l6, yshift = 0.4cm, text width = 6cm,minimum width=5.5cm, draw=magenta] 
{Compute data for denominator in \eqref{avper}\\
${\bf{x}}_\ell \leftarrow {\cal{T}}_L$  };
\draw [arrow] (l6) -- (l7); 
\node (t2) [textt2, left of=l7, xshift=-2.1cm, text width=1cm] 
{7.};

\node (l8) [ifloop, below of=l7, yshift = 0.4cm, text width = 6.2cm,minimum width=4.5cm, draw=magenta,fill=magenta!50] 
{\textbf{If} NTS not available (${\cal{T}}_L = \emptyset$)\\};
\node (t2) [textt2, left of=l8, xshift=-2.1cm, text width=1cm] 
{8.};

\node (l9) [process, below of=l8, yshift = 0.4cm, text width = 6cm,minimum width=4.5cm, draw=magenta] 
{Estimate $\widehat{\boldsymbol{\theta}}_\bn$, the parameters  of model \\ ${\cal{M}}_\bn$  based on the input data ${\bf{x}}$};
\draw [arrow] (l8) -- (l9); 
\node (t2) [textt2, left of=l9, xshift=-2.1cm, text width=1cm] 
{9.};

\node (l10) [process, below of=l9, yshift = 0.2cm, text width = 6cm,minimum width=4.5cm, draw=magenta] 
{Compute data for denominator in \eqref{avper}\\
${\bf{x}}_\ell \leftarrow \widehat{{\bf{\cal{T}}}}_L|{\cal{M}}_\bn(\widehat{\boldsymbol{\theta}}_\bn)$};
\draw [arrow] (l9) -- (l10); 
\node (t2) [textt2, left of=l10, xshift=-2.1cm, text width=1cm] 
{10.};

\node (l11b) [ifloop, below of=l10, yshift = 0.2cm, text width = 6.5cm,minimum width=4.5cm, draw=magenta,fill=magenta!10] 
{\textbf{If} the noise ${\bf{n}}$ is white
};
\draw [arrow] (l10) -- (l11b); 
\node (t2) [textt2, left of=l11b, xshift=-2.1cm, text width=1cm] 
{11.};

\node (l11) [process, below of=l11b, yshift = 0.6cm, text width = 6cm,minimum width=4.5cm, draw=magenta] 
{Estimate the variance $\widehat{\sigma^2}_\bw$ of ${\bf{x}}$};
\draw [arrow] (l11b) -- (l11); 
\node (t2) [textt2, left of=l11, xshift=-2.1cm, text width=1cm] 
{12.};

\node (l12) [process, below of=l11, yshift = 0.4cm, text width = 6cm,minimum width=4.5cm, draw=magenta] 
{Compute data for denominator in \eqref{avper}\\
${\bf{x}}_\ell \leftarrow \widehat{\sigma^2}_\bw~\bf{1}$};
\draw [arrow] (l11) -- (l12); 
\node (t2) [textt2, left of=l12, xshift=-2.1cm, text width=1cm] 
{13.};

\node (l13) [process, below of=l12, yshift = 0.1cm, text width = 6cm,minimum width=4.5cm] 
{Compute ${\overline{\bf p}}_L$ with \textsf{P}, \\ the averaged periodogram of ${\bf{x}}_\ell$};
\draw [arrow] (l12) -- (l13); 
\node (t2) [textt2, left of=l13, xshift=-2.1cm, text width=1cm] 
{14.};

\node (l14) [process, below of=l13, yshift = 0.2cm,  text width = 6cm,minimum width=6cm] 
{Compute ${{\bf{\widetilde{p}}\;|\;{\bf{\overline{p}}}}}_L$, the standardized \\ periodogram with \eqref{eq_Ptilde}};
\draw [arrow] (l13) -- (l14);
\node (t4) [textt2, left of=l14, xshift=-2.1cm, text width=1cm] 
{15.};

\node (l15) [process3, below of=l14, yshift = 0.2cm,  text width = 6cm,minimum width=6cm,yshift=0.2cm] 
{Apply the detection test $\textsf{T}$ to  ${{\bf{\widetilde{p}}\;|\;{\bf{\overline{p}}}}}_L$};
\draw [arrow] (l14) -- (l15); 
\node (t5) [textt2, left of=l15, xshift=-2.1cm, text width=1cm] 
{16.};

\node (out) [startstop, below of=l15, yshift=0.3cm,xshift=0cm, align=left,text width=6cm] 
{OUTPUTS \\
\vspace{0.2cm}
Test statistic $t({\bf{x}})$
};
\draw [arrow] (l15) -- (out);

\end{tikzpicture}
\end{center}
\vspace{-0.5cm}
\caption{{Detection part of the \textsf{3SD} procedure (algorithm 1).}}
\label{fig_algo1} 
\vspace{-0.5cm}
\end{figure}
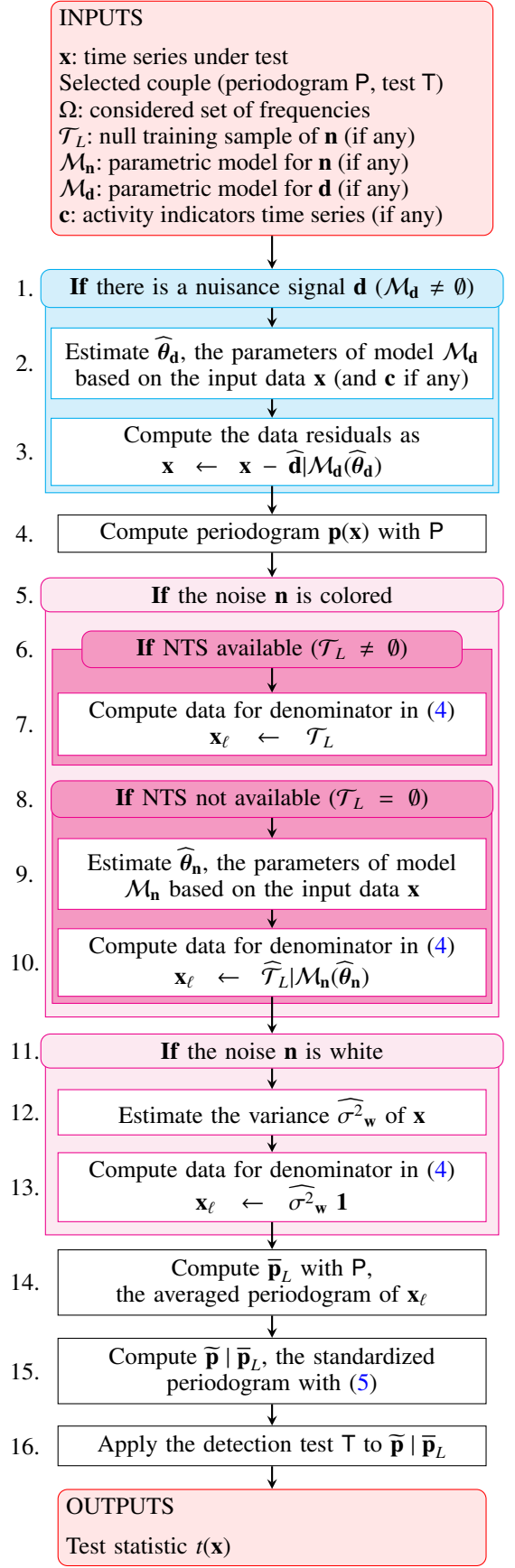

{
The detection procedure proposed in this work is described in Fig.~\ref{fig_algo1}. We refer to this procedure as algorithm~\ref{fig_algo1} in the following.

The first step (rows 1 to 4) aims at computing the periodogram $\bf p$ involved in Eq. \eqref{eq_Ptilde}. 
For this purpose, if the nuisance component $\bd$ is present in model \eqref{hyp}, the algorithm starts by removing it from the time series $\bx$ by fitting  model ${\cal{M}}_\bd$ (blue block). The residuals are used to compute $\bf p$ (row 4).
If $\bd={\bf{0}}$ is not present in \eqref{hyp}, then the {blue block is ignored, and} we directly compute $\bf p$ with the data under test $(\bx)$.

{In this first step, the algorithm removes the nuisance signals, and detection is performed on the residuals. A more general detection approach such as the generalized likelihood ratio  \citep{kaydetection} would require jointly estimating all unknown parameters under  both hypotheses. This approach would lead to different nuisance parameters under ${\mathcal{H}_0}$ and ${\mathcal{H}_1}$ because the parameters of the nuisance signals are consecutively estimated with or without those of the signal (sinusoidal or Keplerian parameters). This procedure can be  computationally very expensive or even intractable, however, when  a maximum likelihood estimation  is required over a  large number of  parameters. It is  therefore discarded in our framework, where nuisance parameters are estimated only under ${\mathcal{H}_0}$. For weak planetary signatures,  the perturbation of these signatures in the nuisance parameters is weak, so that the estimation of these parameters does not change much under both hypotheses.}

The second step (rows 5 to 10) aims at computing the averaged periodogram ${\overline{\bf p}}_L$ involved in Eq. \eqref{eq_Ptilde}.
For this purpose, if $\bn$ in \eqref{hyp} is a colored noise for which an NTS is available (rows 6 to 7), we directly take the $L$ null training series to compute ${\overline{\bf p}}_L$ in row 14.
If no NTS is available (rows 8 to 10),  we fit model ${\cal{M}}_\bn$ to the time series under test $\bx$ and compute a synthetic NTS (called $\widehat{{\bf{\cal{T}}}}_L$) with the estimated model parameters $\widehat{\boldsymbol{\theta}}_\bn$. In this case, ${\overline{\bf p}}_L$ in row 14 is evaluated after generating a sufficiently large number of  synthetic noise time series from $\calM_\bn$.
If noise $\bn$ is white (rows 11 to 13), the averaged periodogram in row 14 simply becomes the product of the estimated variance of $\bx$ ($\widehat{\sigma_{\bw}^2}$) by vector ${{\bf{{1}}}:=[1 \hdots 1]^\top}$.
} 

The final steps of the procedure consist of computing ${{\bf{\widetilde{p}}\;|\;{\bf{\overline{p}}}}}_L$ (row 15) and applying the detection test $\textsf{T}$ (row 16). The procedure returns the test statistics  $t({\bf{x}})$.

\begin{table*}[t]
\centering
\caption{Input parameters used in this work for algorithms~\ref{fig_algo1} and \ref{fig_algo3}, and a nonexhaustive list of additional examples that can be used.}
\label{tabex}
\begin{tabular}{|c|c|c|}
  \hline
  && \\
 Input parameter & Examples selected in this work for illustration & Other examples \\[2ex]
  \hline 
      && \\
  &  &  {Other periodograms' variants:} \\
  &  & e.g., Ferraz-Mello 1981, Cumming et al. 1999, 2004, \\
Periodogram \textsf{P} &  Lomb-Scargle periodogram (Sec.~\ref{sec4},\ref{sec5},\ref{sec6}),  & Reegen 2007, Bourguignon et al. 2007, Baluev 2013, \\
& Generalized Lomb-Scargle  (Sec.~\ref{sec55}) & Jenkins et al. 2014, Tuomi et al. 2014, \\
&  & Gregory, 2016, Hara et al. 2017, 2021 \\[1ex]
 \hline
  && \\
  & Max test $T_M$ (Eq.\eqref{T_Max}), &  Any detection test (see e.g., in Sulis et al. 2017):\\
 Detection test \textsf{T}& Test of the $N_C^{\rm th}$ largest value, $T_C$ (Eq.\eqref{eq_tc}) & e.g., Higher Criticism (Donoho \& Jin 2004),\\
 & & Berk-Jones (Berk \& Jones 1979) \\[1ex]
 \hline
    &&  \\
     & Gaussian process (Eq.\eqref{eq_kernel}), & Any relevant model for $\bd$:  \\
        Noise models $\calM_\bd$ & Linear activity proxy with polynomials (Eq.\eqref{model_a}), &   e.g., models for long term instrumental trends, \\
        & Linear activity proxy (Eq.\eqref{model_b}) & stellar magnetic activity, binary trends \\[1ex]
  \hline
    && \\
   & AR process (Eq.\eqref{eq_AR}) &  Any relevant model for $\bn$:  \\
   Noise models $\calM_\bn$ & Harvey function (Eq.\eqref{eq_harvey}) & e.g., models for meridional circulation, stellar \\
   && granulation, supergranulation, or instrumental noise\\[1ex]
\hline
   && \\
    &  & Any activity proxy for $\bd$: \\
Ancillary time series  $\vect{c}$ & $\log{\rm R'_{HK}}$ (Sec.~\ref{sec5}, Sec.~\ref{sec6}) &  e.g., S-index (Wilson, 1968), FF' (Aigrain et al. 2012), \\
                        &  &  bisector (Queloz et al. 2001) \\[1ex]
  \hline
\end{tabular}
\end{table*}

\subsection{Periodogram and test \sop{used in this work} }
\label{sec23}

In order to illustrate our detection framework, we mainly consider the Lomb-Scargle periodogram here \citepads{1982ApJ...263..835S}. The numerator and the denominator in Eq. \eqref{eq_Ptilde} are therefore LSP.
For the detection test, we consider the  test  whose  statistics is the largest periodogram peak.   This test is referred to as Max test for short. 
The Max test statistic writes
\begin{equation}
 T_{M}({\bf{\widetilde{p}\;|\;\overline{p}}}_L):= \displaystyle{\max_k} ~\widetilde{\bp}(\nu_k).
        \label{T_Max}
\end{equation}

We emphasize that the \textsf{3SD} detection procedure is not tied to the particular choice of the pair (periodogram and test) selected above. It can indeed be applied as such to other types of periodograms and test statistics \sop{that} exploit the periodogram ordinates differently (see \citeads{2016arXiv160107375S} and \citeads{2017ITSP...65.2136S}  for examples of other detection tests). {Table~\ref{tabex} summarizes examples of input parameters that can be used in algorithm~\ref{fig_algo1}}.

{When algorithm~\ref{fig_algo1} has computed the test statistics of $\bx$, the important question is to evaluate its $p$-value}. This is the purpose of Sec.~\ref{sec3}.

\section{{Numerical estimation of the $p$-values} }
\label{sec3}

For an observed  test statistics $t$, the  $p$-value is denoted by $p_{v}(t)$ and defined as 
$$p_{v}(t) := \textrm{Pr} \left(T>t\;|\; {\cal H}_0\right).$$
For example, when all parameters are known, the $p$-value of a test statistics $t$ obtained by the Max test \eqref{T_Max} is defined as
\begin{equation} 
p_{v}(t) := \textrm{Pr} \left(T_M({\bf{\widetilde{p}\;|\;\overline{p}}}_L) > t\;|\; {\cal H}_0\right) = {1- \Phi_{T|{\cal{H}}_0}(t),}
 \label{pfa0}
\end{equation}  
with {$\Phi_{T|{\cal{H}}_0}$} the cumulative distribution function (CDF) of the test statistics under ${{\cal{H}}_0}$.

A $p$-value estimation procedure in which all parameters of the statistical model are known under ${{\cal{H}}_0}$ is called an oracle in the statistical literature (see \citetads{10.2307/2337118}; \citetads{10.1214/21-AOS2141};  \citetads{2021arXiv210613501M} and references therein). We describe such an oracle procedure in Sec.~\ref{sec31}. 

When unknown nuisance parameters are present under {$\mathcal{H}_0$}, the computation of $p$-values is not straightforward.{ A common approach that allows accounting for uncertainties in the model parameters considers the unknown parameters under ${\mathcal{H}}_0$ as random and computes the expectation of $p$-values over the distribution of a prior parameter. 
This is the approach of the Monte Carlo procedure proposed in algorithm~\ref{fig_algo3} (see Sec.~\ref{sec32}). Connections with existing approaches are discussed in Sec. \ref{connec}.} 

In practice, a desirable feature of the procedure is also that it provides a high-probability interval (e.g., 90\%) for the location of the true (i.e., oracle) $p$-value.  The Monte Carlo sampling approach of algorithm~\ref{fig_algo3}  naturally provides estimates of such intervals, called $90\%$ intervals below. As a complementary $p$-value estimate, the highest $p$-value (upper envelope of the sampled $p$-values)  obtained from the simulation can also be used as a (often overly) conservative estimate \citepads{10.2307/2669749}. In practice, a large difference between the two estimates is the sign of a strong dependence of the estimation results on the considered models and parameters.

\subsection{Oracle}
\label{sec31} 

When all parameters under the null hypothesis are known, which is an ideal situation that is not met in practice, a  straightforward way for evaluating the oracle $p$-values numerically consists of performing algorithm~\ref{fig_algo1} many times (e.g., $b$) to obtain a set $\{t({\bf{x}})\}_{j=1,\cdots,b}$ of realizations of the test statistics ${t}$.
{From this set, we can compute an empirical estimate of $\Phi_{T|\mathcal{H}_0}$, called $\widehat{\Phi}_{T}$ for short, and an estimated $p$-value by plugging  $\widehat{\Phi}_{T}$ into Eq. \eqref{pfa0}. The estimation of the $p$-value  can be made arbitrarily accurate because  $\widehat{\Phi}_{T}$ converges to $\Phi_{T|\mathcal{H}_0}$  as $b\to \infty$.}

\subsection{Algorithm for estimating the $p$-values } 
\label{sec32}

The generic procedure (algorithm~\ref{fig_algo3}) is presented in Fig.~\ref{fig_algo3}.  The final expected $p$-value ($\overline{\widehat{p}_v}$) is obtained by sample expectation (row 19) over $B$ $p$-values $\widehat{p}_v^{\{i\}}$ obtained in a major loop (green, rows 1 to 18). Each such $p$-value is obtained through a minor loop (rows 5 to 16) that measures an empirical CDF ($ \widehat{\Phi}^{\{i\}}_{T}$). {This loop (and this estimated CDF) includes an averaging over the priors $\pi_{\bd}$  and  $\pi_{\bw}$. }

The major loop samples $B$ values of the parameters for $\bn$ ($\widehat{\widehat{\btheta}}_\bn^{\{i\}}$), which are obtained by parametric bootstrap: a sample NTS is generated according to parameters ${\widehat{\btheta}}_\bn$ estimated from the data (rows 3 and 4). 
In the nested minor loop (blue), the algorithm samples a number $b$ of test statistics $t$ produced by algorithm~\ref{fig_algo1} conditionally  on $\widehat{\widehat{\btheta}}_\bn^{\{i\}}$ (if $\bn$ is colored), $\widehat{\widehat{\sigma^2}}_\bw ^{\{i,j\}}$ (if $\bn$ is white) and on $\widehat{\widehat{\btheta}}_\bd^{\{i,j\}}\!$ (if $\bd\neq {\bf{0}}$ in Eq. \eqref{hyp}). These parameter values are obtained by randomly perturbing the corresponding input values $\widehat{\sigma^2}_\bw$ and $\widehat{\btheta}_\bd|{\cal{M}}_\bd$ within reasonable uncertainties according to some prior distributions, called $\pi_\bw$ (row 10) and $\pi_{\bd}$ (row 13). 
If model ${\cal{M}}_\bd$ involves the use of an ancillary series $\vect{c}$, a synthetic series $\widehat{\vect{c}}^{\{i,j\}}$ is generated in row 14 using the same model parameters $\widehat{\widehat{\btheta}}_\bd^{\{i,j\}}\!$ plus a WGN.

Each test statistics is then obtained by applying algorithm~\ref{fig_algo1} to simulated time series $\bx^{\{i,j\}}$ (row 16). If $\bn$ is colored, algorithm~\ref{fig_algo1} takes as input the $L$ simulated NTS ${\mathcal{T}}_L^{\{i,j\}}$ (computed in row 7) for periodogram standardization. If $\bn$ is white, this vector is sent as empty in algorithm~\ref{fig_algo1}.

{A particular case of this algorithm arises when the noise is considered colored with a parametric model ${\cal{M}}_\bn$ and no NTS is available. This case impacts the setting of $L$ in rows 3 and 7 and is explained in Sec. \ref{sec55}}.

For $\pi_\bw$ and $\pi_{\bd}$, we consider Gaussian and uniform priors (with independent components if multivariate) in the simulations below. The component distributions are centered on the parameter values estimated from the data. The scale parameters (widths of the intervals for uniform priors or variances for the Gaussian prior) are estimated from the estimation error bars.

\subsection{{Connection with the literature}}
\label{connec}

\begin{figure*}[t!]
\centering
\resizebox{\textwidth}{!}{%
\tikzstyle{startstop} = [rectangle, rounded corners, minimum width=2cm, minimum height=0.5cm,text centered, text width=7cm, draw=red, fill=red!10]

\tikzstyle{ifloop} = [rectangle, rounded corners, minimum width=2cm, minimum height=0.5cm,text centered, text width=7cm, draw=cyan, fill=white]

\tikzstyle{endstop} = [rectangle, rounded corners, minimum width=\textwidth, minimum height=0.5cm,text centered, text width=\textwidth, draw=red, fill=white]

\tikzstyle{endstop2} = [rectangle, rounded corners, minimum width=0.4*\textwidth, minimum height=0.5cm,text centered, text width=\textwidth, draw=red, fill=white]

\tikzstyle{process} = [rectangle, minimum width=3cm, minimum height=0.5cm, text centered, text width=5.5cm, draw=black, fill=white]

\tikzstyle{process1} = [rectangle, minimum width=1cm, minimum height=0.5cm, text centered, text width=2cm, draw=black, fill=white]

\tikzstyle{process2} = [rectangle, minimum width=2cm, minimum height=0.5cm, text centered, text width=4cm, draw=black, fill=white]

\tikzstyle{process3} = [rectangle, minimum width=2cm, minimum height=0.5cm, text centered, text width=3.5cm, draw=black, fill=white]

\tikzstyle{scheme} = [rectangle, minimum width=3cm, minimum height=0.5cm, text centered, text width=3cm, draw=black, fill=white]

\tikzstyle{losange} = [diamond, minimum width=2cm, minimum height=0.5cm, text centered, draw=black, fill=white]

\tikzstyle{textt} = [rectangle, minimum width=5cm, minimum height=0.5cm, text centered, text width=6cm, draw=white, fill=white]

\tikzstyle{textt2} = [rectangle, minimum width=5cm, minimum height=0.5cm, text centered, text width=0.5cm]

\tikzstyle{arrow} = [thick,->,>=stealth]
\tikzstyle{line} = [thick,-,>=stealth]

\tikzstyle{MCloop} = [thick,-,>=stealth]

\tikzstyle{header} = [rectangle, minimum width=\textwidth, minimum height=0.5cm, text centered, text width=\textwidth, draw=black, fill=white]

\definecolor{ao(english)}{rgb}{0.0, 0.5, 0.0}
\definecolor{blue(pigment)}{rgb}{0.2, 0.2, 0.6}

\hspace{-10cm}
\begin{tikzpicture}[node distance=1.4cm] 



\node (b0) [textt, text width=25cm] {};

\node (start) [startstop,below of=b0, yshift = 0.1cm, align=justify]
{
INPUTS \\ 
\vspace{-0.2cm}\\
${\bf x}$: time series under test\\
Selected couple (periodogram $\textsf{P}$, test $\textsf{T}$) \\
$\Omega$: considered set of frequencies\\
$B, b$: number of Monte Carlo simulations\\
\\
${\cal{M}}_\bn$: parametric model for ${\bf{n}}$ (if any) \\
$\widehat{\btheta}_\bn|{\cal{M}}_\bn$: estimated parameters (if any)  \\
\\
$\widehat{\sigma^2}_\bw$: estimated variance of WGN (if any)\\
$\pi_\bw$: parameters' prior distribution for $\widehat{\sigma^2}_\bw$ with scale parameter  $\widehat{\bdelta_\bw}$ (if any)\\
\\
${\cal{M}}_\bd$: parametric model for ${\bf{d}}$ (if any) \\
$\widehat{\btheta}_\bd|{\cal{M}}_\bd$: estimated parameters (if any)\\
$\pi_\bd$: parameters' prior distribution for $\widehat{\btheta}_\bd$ with scale parameter  $\widehat{\bdelta_\bd}$ (if any) \\
${\bf c}$: ancillary time series (if any)
};

\draw [arrow,draw=red]  (3.7,2) -- (4.7,2);

\draw [green,fill=green!10] (4.8,-18) -- (14,-18) -- (14,2.8) -- (4.8,2.8)-- cycle;
\draw [magenta,fill=magenta!10]   (5.1,-0.7) -- (13.8,-0.7) -- (13.8,1.8) -- (5.1,1.8)-- cycle;
\draw [blue,fill=blue!10]   (5.1,-1.4) -- (13.8,-1.4) -- (13.8,-15.8) -- (5.1,-15.8)-- cycle;
\draw [magenta,fill=magenta!10] (5.7,-2.3) -- (13.1,-2.3) -- (13.1,-5.3) -- (5.7,-5.3)-- cycle;
\draw [magenta,fill=magenta!10] (5.7,-6) -- (13.1,-6) -- (13.1,-9.1) -- (5.7,-9.1)-- cycle;
\draw [cyan,fill=cyan!15]   (5.7,-14.3) -- (13.1,-14.3) -- (13.1,-9.7) -- (5.7,-9.7)-- cycle;

\node (l1) [process2, right of=start, draw=green,xshift=8cm,yshift=4.1cm, text width = 6cm,minimum width=6cm, fill=green!10] 
{\textcolor{ao(english)}{Compute a set of $i = \{1,\cdots,B\}$ $p$-values} };
\node [textt2, right of=l1, xshift=3.5cm, text width=1cm] 
{1.};

\node (l2) [ifloop, below of=l1, yshift = 0.4cm, text width = 7.5cm,minimum width=5.5cm, draw=magenta,fill=magenta!10]
{\textbf{If} the  noise ${\bf{n}}$ is colored};
\node [textt2, right of=l2, xshift=3.5cm, text width=1cm] 
{2.};

\node (l3) [process, below of=l2, draw=magenta,yshift=0.4cm, text width = 8cm] 
{Generate $L$ noise time series ${\cal{T}}^{\{i\}}$ with model ${\cal{M}}_\bn(\widehat{\btheta}_\bn)$};
\draw [arrow] (l2) -- (l3); 
\node [textt2, right of=l3, xshift=3.5cm, text width=1cm] 
{3.};

\node (l4) [process, below of=l3, draw=magenta,yshift=0.4cm, text width = 8cm] 
{Fit ${\cal{T}}^{\{i\}}|{\cal{M}}_\bn$ and generate new estimates $\widehat{\widehat{\btheta}}_\bn^{\{i\}}$};
\draw [arrow] (l3) -- (l4); 
\node [textt2, right of=l4, xshift=3.5cm, text width=1cm] 
{4.};

\node (l5) [process2, below of=l4, draw=blue,yshift=0.2cm, text width = 7cm,minimum width=6cm,  fill=blue!10] 
{\textcolor{blue(pigment)}{Compute a set of $j = \{1,\cdots,b\}$ test statistics }};
\node [textt2, right of=l5, xshift=3.5cm, text width=1cm] 
{5.};

\node (l6) [ifloop,  below of=l5, draw=magenta,yshift=0.5cm, text width = 7.5cm,minimum width=4.5cm, draw=magenta,fill=magenta!10] 
{\textbf{If} the  noise ${\bf{n}}$ is colored};
\node  [textt2, right of=l6, xshift=3.5cm, text width=1cm] 
{6.};

\node (l7) [process, below of=l6, draw=magenta,yshift=0.3cm, text width = 7cm] 
{Generate $L+1$ training series ${\cal{T}}_{L+1}^{\{i,j\}}$ with ${\cal{M}}_\bn(\widehat{\widehat{\btheta}}_\bn^{\{i\}})$};
\draw [arrow] (l6) -- (l7); 
\node  [textt2, right of=l7, xshift=3.5cm, text width=1cm] 
{7.};

\node (l8) [process, below of=l7, draw=magenta,yshift=0.1cm, text width = 7cm] 
{Initialize the synthetic time series under test\\
with one series: ${\bf{x}}^{\{i,j\}} \leftarrow {\cal{T}}_1^{\{i,j\}}$};
\draw [arrow] (l7) -- (l8); 
\node  [textt2, right of=l8, xshift=3.5cm, text width=1cm] 
{8.};

\node (l9) [ifloop, below of=l8, draw=magenta,yshift=0.2cm, text width = 7.5cm,minimum width=4.5cm,fill=magenta!10] 
{\textbf{If} the  noise ${\bf{n}}$ is white};
\node  [textt2, right of=l9, xshift=3.5cm, text width=1cm] 
{9.};

\node (l10) [process, below of=l9, draw=magenta,yshift=0.3cm, text width = 7cm] 
{Sample $\widehat{\widehat{\sigma^2}}_\bw ^{\{i,j\}}$ from prior $\pi_\bw$ };
\draw [arrow] (l9) -- (l10); 
\node  [textt2, right of=l10, xshift=3.5cm, text width=1cm] 
{10.};

\node (l11) [process, below of=l10, draw=magenta,yshift=0.0cm, text width = 7cm] 
{Initialize the synthetic time series under test\\
${\bf{x}}^{\{i,j\}} \leftarrow \boldsymbol{w}^{\{i,j\}}\!\sim\!{\cal{N}}({\bf{0}},\widehat{\widehat{\sigma^2}}_\bw ^{\{i,j\}}\bf{I} )$};
\draw [arrow] (l10) -- (l11); 
\node  [textt2, right of=l11, xshift=3.5cm, text width=1cm] 
{11.};

\node (l13) [ifloop, below of=l11, yshift = 0.2cm, text width = 7.5cm,minimum width=4.5cm,  draw=cyan,fill=cyan!15] 
{\textbf{If} there is a nuisance signal ${\bf{d}}$ (${\cal{M}}_\bd \neq \emptyset$)};
\node  [textt2, right of=l13, xshift=3.5cm, text width=1cm] 
{12.};

\node (l14) [process, below of=l13, draw=cyan,yshift=0.3cm, text width = 7cm] 
{Sample $\widehat{\widehat{\btheta}}_\bd^{\{i,j\}}\!$  from prior $\pi_\bd$};
\draw [arrow] (l13) -- (l14); 
\node  [textt2, right of=l14, xshift=3.5cm, text width=1cm] 
{13.};

\node (l14b) [process, below of=l14, draw=cyan,yshift=-0.1cm, text width = 7cm] 
{Generate a synthetic ${\bf{d}}$ series: $\widehat{\bd}^{\{i,j\}}\Big|{\cal{M}}_\bd\Big(\widehat{\widehat{\btheta}}_\bd^{\{i,j\}}\!\!\!, \widehat{\vect{c}}^{\{i,j\}}\Big)$\\
with $\widehat{\vect{c}}^{\{i,j\}}$ a synthetic ancillary series (if ${\bf c} \neq \emptyset$)  \\};
\draw [arrow] (l14) -- (l14b); 
\node  [textt2, right of=l14b, xshift=3.5cm, text width=1cm] 
{14.};

\node (l15) [process, below of=l14b, draw=cyan,yshift=-0.1cm, text width = 7cm] 
{Add it to the synthetic time series under test\\
${\bf{x}}^{\{i,j\}} \leftarrow {\bf{x}}^{\{i,j\}} + \widehat{\bd}^{\{i,j\}}$};
\draw [arrow] (l14b) -- (l15); 
\node  [textt2, right of=l15, xshift=3.5cm, text width=1cm] 
{15.};

\node (l17) [process, below of=l15, draw=blue,yshift=-0.cm, text width = 7.5cm] 
{Compute test statistic of ${\bf{x}}^{\{i,j\}}$:\\
$t({\bf{x}}^{\{i,j\}}) \!=\!\! {\textrm{algorithm}\ref{fig_algo1}}\Big({\bf{x}}^{\{i,j\}},(\textsf{P},\textsf{T}), \bOmega, {\cal{T}}_L^{\{i,j\}},{\cal{M}}_\bd, \widehat{\vect{c}}^{\{i,j\}} \Big)$};
\node  [textt2, right of=l17, xshift=3.5cm, text width=1cm] 
{16.};

\node (l18) [process, below of=l17, draw=green,yshift=0.1cm, text width = 8cm] 
{Compute the test CDF $\widehat{\Phi}_{T}^{\{i\}}(t)$ using the sample  $t({\bf{x}}^{\{i,j\}})$};
\node  [textt2, right of=l18, xshift=3.5cm, text width=1cm] 
{17.};

\node (l19) [process, below of=l18, draw=green,yshift=0.3cm, text width = 8cm] 
{Compute the $p$-value $\widehat{p}_{v}^{\{i\}}(t)= 1 - \widehat{\Phi}_{T}^{\{i\}}(t)$};  
\draw [arrow] (l18) -- (l19); 
\node  [textt2, right of=l19, xshift=3.5cm, text width=1cm] 
{18.};

\node (l20) [process, below of=l19, draw=black,yshift=0.2cm, text width = 8.9cm] 
{Compute the mean $p$-value $\overline{\widehat{p}}_{v}(t) = \frac{1}{B} \sum_{i=1}^{B} \widehat{p}_{v}^{\{i\}}(t)$ };
\draw [arrow] (l19) -- (l20); 
\node  [textt2, right of=l20, xshift=3.5cm, text width=1cm] 
{19.};


\node (DPFA) [startstop, left of=l20, yshift=0.2cm,xshift=-8cm,align=left, text width = 7.5cm] 
{OUTPUTS \\
\vspace{0.2cm}
$\overline{\widehat{p}}_{v}(t)$: estimated mean $p$-value and credible interval};
\draw [arrow,draw=red] (4.7,-18.9) -- (3.9,-18.9);

\end{tikzpicture}
}
\caption{{Part of the \textsf{3SD} procedure that estimates the $p$-value of algorithm~\ref{fig_algo1} (algorithm 2).}}
\label{fig_algo3}  
\end{figure*}
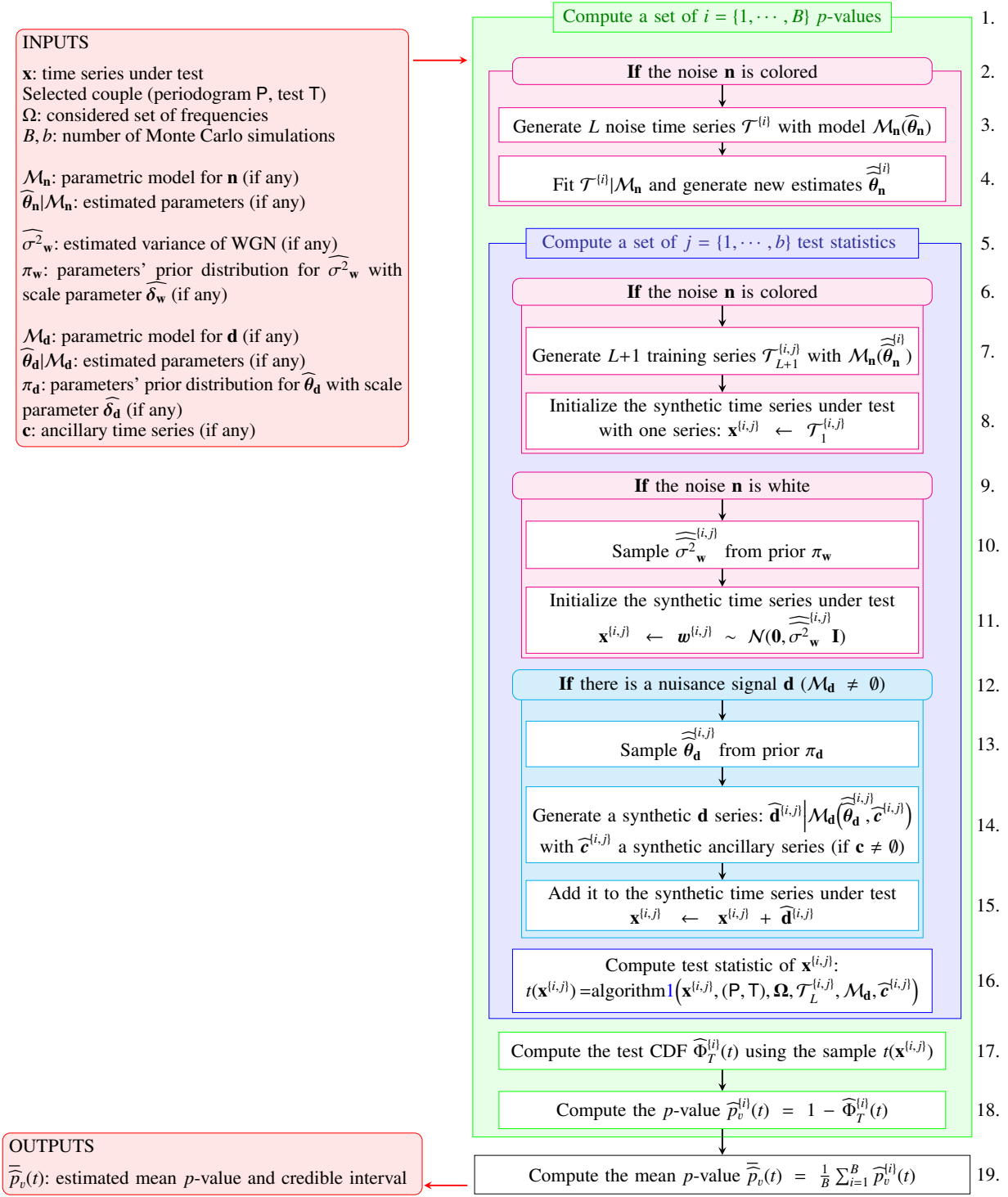

{ In the statistical literature, several approaches exist for estimating $p$-values when unknown parameters are involved in the detection process (see \citeads{10.2307/2669749} for a review). In these approaches, prior predictive $p$-values are obtained by averaging over a prior  law of the parameters \citep{10.2307/2982063}. In contrast, posterior predictive $p$-values  require averaging over the posterior distribution \citep{guttman1967use,rubin1984bayesianly}. Sampling from the posterior would require a computationally involved procedure (e.g., with a Markov chain Monte Carlo method). For this reason, we do not follow the posterior predictive approach here and aim at keeping the sampling process simple. In our approach, this process is different depending on the parameters we consider. We list the different cases below.
\begin{itemize}
    \item  For the parameters of ${\cal{M}}_n$ (row 4 of algorithm \ref{fig_algo3}), which correspond to a parametric model of a random process, a natural way of sampling parameters over a distribution that is consistent with our model is to sample the random process with the estimate at hand and to re-estimate the parameters. This is what we do here. This approach also means that we do not have to choose an  explicit prior for ${\boldsymbol{\theta_n}}$,  the parameters of which would need to be set or estimated.\\
    \item For the other unknown parameters $\sigma_{\bw}$ and ${\boldsymbol{\theta_d}}$ in rows 10 and 13 of algorithm \ref{fig_algo3}, we did not use the same strategy for computational reasons. Here, we found it more efficient to sample these  parameters from some priors (Gaussian and uniform), whose parameters were estimated from the estimated error bars in an empirical Bayes manner  \citep{efron_hastie_2016}.
\end{itemize}
The performances of the generic algorithm \ref{fig_algo3} are evaluated on real and synthetic data in the next three sections.}\\

\section{Noise is granulation and WGN} 
\label{sec4}

\subsection{{Objectives}}
\label{sec41}

We first consider the simplest case in which the RV time series is only affected by the{ stochastic} colored noise $\bn$ in Eq. \eqref{hyp}, of which a training set $\calT_L$ is assumed to be available (i.e., we set $\bd={\bf{0}}$ in Eq.  \eqref{hyp}, and ${\cal{M}}_\bd=\emptyset$ in algorithms~\ref{fig_algo1} and \ref{fig_algo3}).
{While this configuration is not realistic for most long-period planet RV searches, where magnetic activity phenomena (defined in $\bd$) always play an important role in the observed RV time series, it can be useful for detecting low-mass ultra-short period planets (USP, defined with periods $< 1$ day). This has been discussed in \citetads{2020A&A...635A.146S}. In addition to this very specific case, this section has two goals.} 
The first goal is pedagogical because we  illustrate the principle of the detection procedure  with the simplest case. 
The second goal is to validate the proposed $p$-value estimation procedure when the NTS is composed with MHD simulations of stellar granulation and for irregular sampling. So far, this validation has only been performed for a regular sampling, in which case the FAP and $p$-values can be computed with analytical expressions  \citepads{2020A&A...635A.146S}.

\subsection{Validation of the \textsf{3SD} procedure on solar data}
\label{sec42}

\subsubsection{Data, MHD simulations, and sampling grids}
\label{sec421}

{The validation of the procedure on real data poses two difficulties. 
First,  establishing a ground-truth reference for the $p$-value that would be obtained with an NTS composed of genuine stellar convection noise requires implementing the oracle procedure (see Sec.~\ref{sec31}), which in turn requires a very large number of genuine stellar noise time series. 
Fortunately, the GOLF/SoHO spectrophotometer has made $25$-year-long time series of solar RV observations available that can be used (e.g., see \citeads{2018A&A...617A.108A}).
Second, in this section, we focus on granulation and instrumental noise. We therefore need to disregard the low-frequency part of the noise PSD that is dominated by the magnetic activity. For this purpose, the detection test is applied to frequencies $\nu > 56$ $\mu$Hz alone because {at frequencies below $56$ $\mu$Hz,} the PSDs of solar data and MHD simulations start diverging (see \citeads{2020A&A...635A.146S} and Appendix~\ref{appB}). }

 Because GOLF observations cover a very long period of time and because the timescales characteristic for granulation are shorter than a day, we divided the total time series into smaller series and focused on this short-timescale solar variability alone. 
We selected $b=1420$ two-day-long series from the $\approx$ $3000$ available time series taken during the first $17$ years of GOLF observations (the last years are particularly affected by detector aging).

These time series are regularly sampled ($N_{reg}=2880$ data points) at a sampling rate $\Delta t=60$s. As detailed in \citetads{2020A&A...635A.146S}, we filtered out the high-frequency p-modes on each two-day-long series (with periods $<15$ min because they are not reproduced by the MHD simulations), and we added a WGN with variance $\widehat{\sigma}^2_{\bw}$ estimated from the high-frequency plateau visible in the periodogram of the GOLF data (see Appendix~\ref{appB}). The resulting dataset constitutes a large set of noise time series from which the oracle reference results can be computed.

For the implementation of algorithm~\ref{fig_algo3}, the  available  NTS is composed with a $53$-day-long noise time series obtained from  3D MHD simulations of solar granulation (Bigot et al., in prep.). After splitting them into two-day-long time series, $26$ granulation noise time series are available, of which we randomly selected $L=5$ for the standardization.
Before running algorithm~\ref{fig_algo3}, we applied the same corrections as for the GOLF data  to the NTS obtained from MHD simulations ($\calT_5$)  (i.e., filtering of the stellar acoustic modes and addition of an instrumental WGN). The NTS was finally sampled as the observations under test (see the three sampling grids detailed below).

Algorithm~\ref{fig_algo3} also relies on parameters that set the size of the Monte Carlo simulations, $B$ and $b$ (taken here as $B=100$ and  $b=10^4$) and on model ${\cal{M}}_\bn$ with associated parameters $\widehat{\boldsymbol{\theta}}_\bn$. We chose an AR model because it suits our MHD NTS time series well and its parameters are fast to estimate. An AR  process is defined as \citepads{Brockwell_1991} 
\begin{equation}
  \label{eq_AR}
\bn(t_j) = \displaystyle{\sum_{j=1}^{o_{AR}}} \alpha_{AR,j} ~ \bn(t_j - j\Delta t) + {\bf{w}}(t_j),~~~~~~j=1,\hdots,N
,\end{equation}
with $\bn(t_j)$  the sample of $\bn$ at time instant $t_j$, $\alpha_{AR,j}\in~]\!-\! 1; 1[$ the AR coefficients filter, $o_{AR}$ the AR order, $\Delta t$ the sampling time step, and  ${\bf{w}}(t_j)$ the sample at time $t_j$ of a WGN series $\bf{w}$ of variance $\sigma_{\bw}^2$. 
The model parameters $\widehat{\boldsymbol{\theta}}_\bn$ are the AR order and the filter coefficients.  The order was selected using Akaike's final prediction error criterion  \citep{Akaike_1969} and led to a value of $10$, which remained fixed for the remaining procedure.  The filter coefficients (estimated $B\times b +1$ times) were obtained through the standard Yule-Walker equations \citep{Brockwell_1991}. 

For this study, we considered three temporal sampling grids. 
First, we considered a regular sampling grid ($\Delta t = 10$ minutes), where we keep only $10\%$ of the initial GOLF series to make the number of data points $N$ that are comparable to ordinary RV time series. Second, we considered a regular sampling grid ($\Delta t=60$ seconds) affected by a large central gap: two 2.4 h continuous observations separated by a 43.2 h gap. Third, we considered a random sampling grid, in which $10\%$ of the data points are randomly selected in the initial regular grid of the GOLF series ($\Delta t \in [60, 2880]$ seconds). In each case, the sampling grid contains $N=288$ data points. For the regular (irregular) sampling grids, the standardized periodogram was computed using the classical (Lomb-Scargle) periodogram. They are the same for regular sampling. 

\begin{figure*}[t!]
\resizebox{\hsize}{!}{\includegraphics{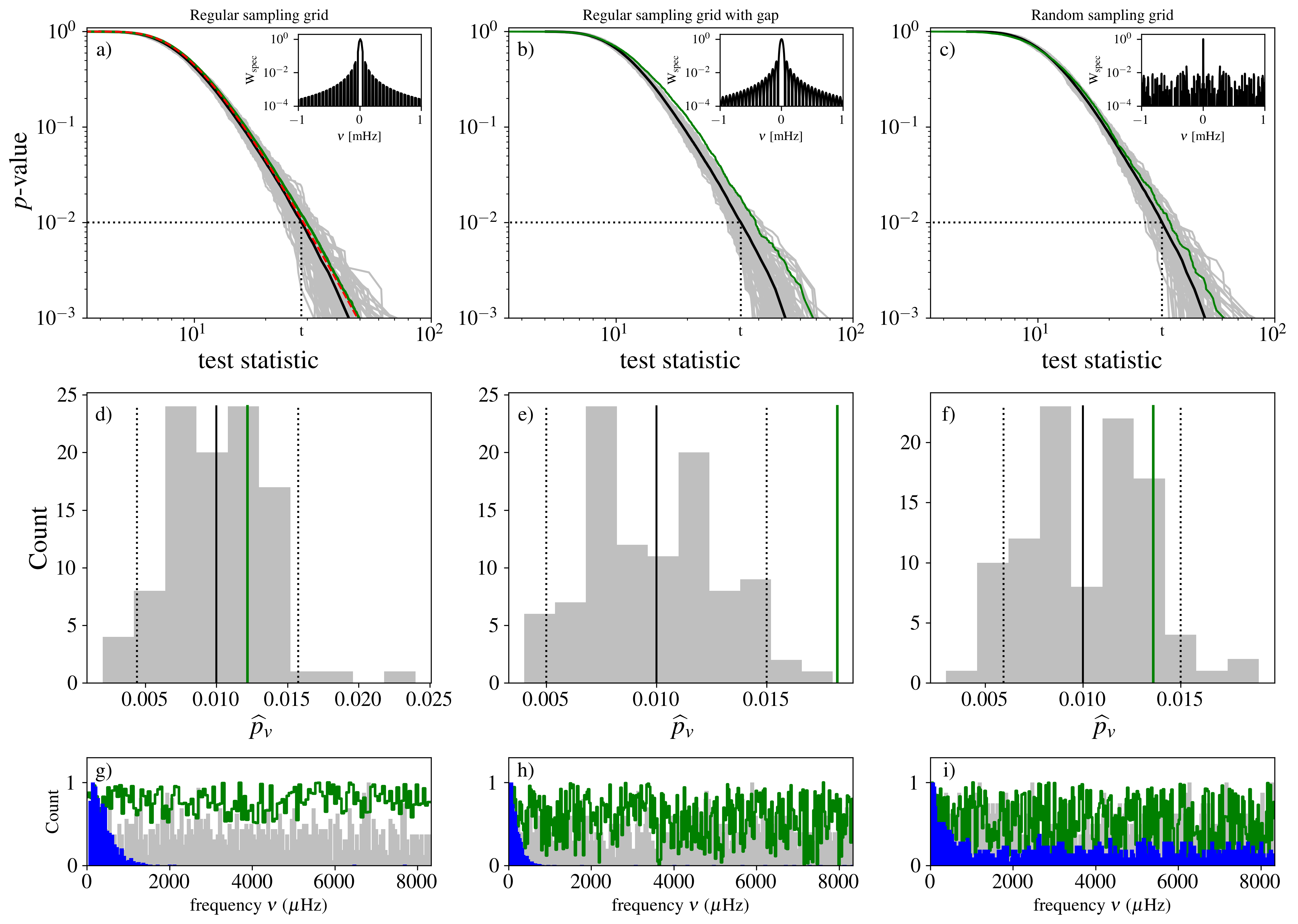}}
\caption{
Validation of the $\textsf{3SD}$ approach for a noise $\bn$ composed of granulation plus WGN and $\bd = {\bf{0}}$ in Eq. \eqref{hyp}. We investigate three temporal sampling grids, from left to right: regular, regular with a central gap, and random.
 \textbf{Top panels:} Relation $p_{v}(t)$ for the detection procedure using a genuine NTS based on GOLF solar data (green; oracle).  The  $B=100$ estimates $\widehat{p}^{\{i\}}_{v}$  produced by the \textsf{3SD} procedure using the MHD-based NTS (gray). Sample mean value $\overline{\widehat{p}}_{v}$ of the gray curves (black). The spectral window of each sampling grid is shown in the inset panels. In panel {a)}, the red line represents the analytical expression of $p_{v}$ (see Eq.(13) of \citeads{2020A&A...635A.146S}). The dotted lines indicate the test statistics for which $\overline{\widehat{p}}_{v}$ equals a mean $p$-value of $1\%$.
\textbf{Middle panels:} Empirical distribution of $\widehat{p}^{\{i\}}_{v}$ at $t$ and $90\%$ intervals (dotted lines) around the mean (solid black lines). {$P$-values  obtained with the oracle are shown with green vertical lines.}
\textbf{Bottom panels:} For one of the $B$ loops of algorithm~\ref{fig_algo3}, we show an example of the empirical distribution of the $b=10^4$ random values of the frequency index where the largest periodogram peak was found (gray). Empirical distribution of the Max test related to the oracle procedure (green; only contours are shown for clarity). The same distribution, but for a nonstandardized LSP is shown in blue (i.e., the Max test is applied to $\bf p$ instead of ${{\bf{\widetilde{p}}\;|\;{\bf{\overline{p}}}}}_L$).
 }
\label{fig2}
\end{figure*}

\subsubsection{{Validation of algorithm 2 on solar data}}
\label{sec422}

The $p$-value estimation procedure provided by algorithm~\ref{fig_algo3} results in  $B=100$ estimates  $\widehat{p}^{\{i\}}_{v}$  shown in gray in the top panels of Fig.~\ref{fig2}. The average $p$-value $\overline{\widehat{p}}_{v}$ is shown in black and the $p$-value of the oracle procedure based on real stellar noise in green. 
\sop{To compute this oracle, we ran algorithm~\ref{fig_algo1} with $1420$ GOLF time series that we further permuted $\text{ten}$ times by blocks.} This led to an effective number of $b=14 200$ series. To compute each realization of the test, we randomly selected one series to compute the periodogram in row 4 of algorithm~\ref{fig_algo1} and $L=5$  other series to compute the averaged periodogram in row 14. 

The beam of the gray curves represents particular estimates of the unknown relation $p_{v}(t)$  of algorithm~\ref{fig_algo1}, with parameters that are consistent with the NTS at hand. The true $p$-value of this procedure is unknown (for irregular sampling) because evaluating it exactly would require an infinite number of MHD simulations.
If the simulations performed in algorithm~\ref{fig_algo3} sample the space of parameters correctly, the truth is likely to lie somewhere in these curves, perhaps close to the mean where the bunch of curves is located.\\

A comparison of the black and green curves shows that for the three considered sampling grids the approach based on MHD simulations is equivalent to using true stellar noise on average. 

This shows that while algorithm~\ref{fig_algo3} only has a reduced number ($L=5$) of synthetic time series for the NTS at hand, it can generate a beam of situations whose mean succeeds in evaluating the true $p$-value of a standardized detection procedure supervised by genuine stellar noise.\\ 
We note a slight mismatch in the middle and right panel that probably arises because our ground-truth (GOLF) data contain some effects that were not accounted for in the $p$-value estimation procedure. Here, the highest $p$-value (upper envelope of the gray curves) provides a more conservative and more reliable estimate than the mean. 
The test statistics $t$ corresponding to a mean $p$-value of $1\%$ are $28.3, 33.6, \text{and }33.5$ for the three samplings. For these values, the simulations show that the $p$-value has an empirical probability of $90 \%$ to lie in the range [$ 0.4\%, 1.6\%$], [$0.5\%, 1.5\%$], and [$0.6\%, 1.5\%$] (panels in the middle row of Fig.~\ref{fig2}). At these $t$ values, the oracle $p$-values correspond to $1.2\%, 1.8\%, \text{and }1.3\%$. 

Because MHD simulations are made to represent stellar granulation noise with high accuracy, the good match observed in each case between the mean $p$-value obtained by algorithm~\ref{fig_algo3} and the oracle based on true stellar noise suggests that the true $p$-value is close to the sample average. 
We also note that in the case of regular sampling, for which the $p$-value expression is known and given in Eq. (13) of \citetads{2020A&A...635A.146S}, theory, the averaged estimate from algorithm~\ref{fig_algo3}, and the oracle with true stellar noise (panel a) match well.

Finally, the bottom panels show the histograms of the distribution of the frequency index in which the  maximum of the standardized periodogram was found during the iterations of algorithm~\ref{fig_algo3} of the \textsf{3SD} procedure (gray) and oracle (green).  
We show the same distribution, but for the (nonstandardized) LSP in blue, that is, a procedure that would apply the Max test to ${\bf{p}}$ instead of $\widetilde{\bf{p}}$ in algorithm~\ref{fig_algo1}. Because the noise is colored, the LSP tends to peak at lower frequencies, where the PSD has higher values (see Fig.~\ref{fig0}).
Colored noise inexorably causes an increased rate of false alarms at these frequencies, but this effect cannot be quantified or even suspected by the user when the noise PSD is unknown. In contrast, the  standardization used in Eq. \eqref{eq_Ptilde} leads to an autocalibration of the periodogram ordinates, which  leads to a more reasonable (uniform) distribution of the frequency index of the maximum peak. 
These results show that the $\textsf{3SD}$ procedure is well calibrated and algorithm~\ref{fig_algo3} succeeds in estimating the resulting $p$-value correctly.  Because the distribution of the test statistics is very similar for data that are standardized with true noise and for data that are standardized by means of MHD noise time series, these results also demonstrate that the MHD simulations can efficiently be exploited in place of a real noise NTS for a periodogram standardization.

\begin{figure}[t]
\resizebox{\hsize}{!}{\includegraphics{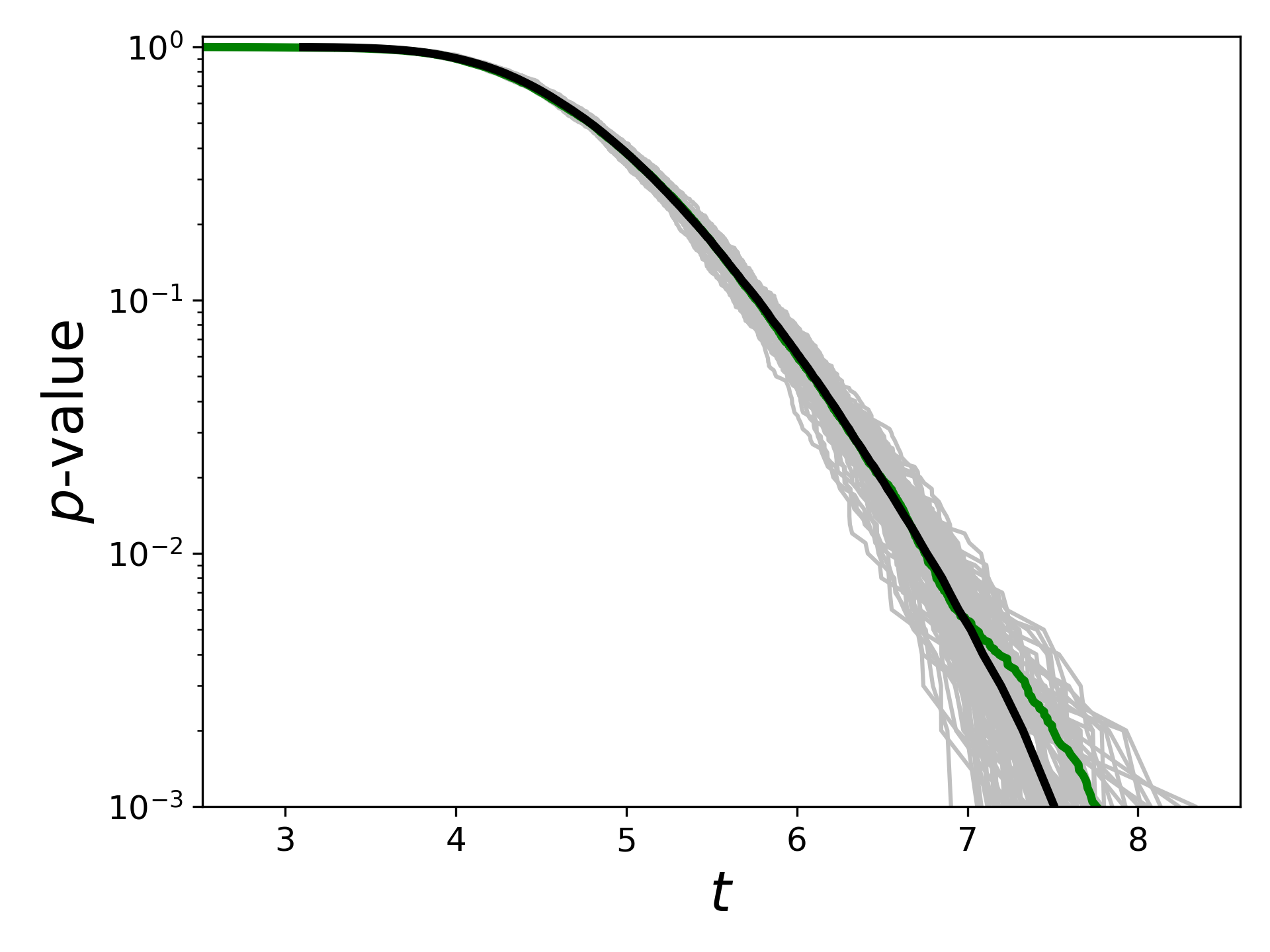}}
\caption{Validation of algorithm~\ref{fig_algo3} using the GLS periodogram and test $T_C$ in Eq. \eqref{eq_tc}. Oracle is shown in green, $p$-value estimates from algorithm~\ref{fig_algo3} are shown in gray, and their mean value is plotted in black. } 
\label{fig3b}
\end{figure}

Finally, we carried out a similar validation of the output $p$-values, but for another pair (periodogram and test). We selected the GLS periodogram \citepads{2009A&A...496..577Z} and the test statistics of the $\rm N_C^{\text{th}}$ maximum periodogram value,  defined as
\begin{equation}
 T_{C}({\bf{\widetilde{p}\;|\;\overline{p}}}_L):= {\widetilde{p}_{({N}-N_C+1)}},
        \label{eq_tc}
\end{equation}
with $N_C\in \mathbb{N}^+$ a parameter related to the number of periodic components, and $\widetilde{p}_{(k)}$ the kth-order statistic of ${\bf{\widetilde{p}}}$.
We note that test $ T_{C}$ may be more discriminative against the null than the Max test $\displaystyle{\max_k} ~\widetilde{p}(\nu_k) = {\widetilde{p}_{({N})}}$ in Eq. \eqref{T_Max} for multiple periodic components, as discussed by \citet{https://doi.org/10.1111/j.2517-6161.1989.tb01762.x} and \citetads{1971GeoJ...23..373S}.
The $p$-values resulting from algorithm~\ref{fig_algo3} computed for $N_C=10$ in Eq. \eqref{eq_tc} and for the random sampling grid defined in Sec. \ref{sec421} are shown in Fig.~\ref{fig3b}. These results illustrate that the reliability of the procedure is not tied to the pair (LSP and Max test).

\section{Noise is magnetic activity, long term trends, granulation and WGN} 
\label{sec5}

\subsection{Setting and objectives}
\label{sec51}

In practice, many stellar activity processes affect the RV observations \citepads{2021arXiv210406072M}, and an NTS is not  available for all of them. This is the case, for example, of stellar magnetic activity such as starspots and plages. We present  a more general application of the \textsf{3SD}  procedure, in which classical techniques that are used to correct for the stellar variability of magnetic sources are coupled with the use of an NTS.  We also investigate the case in which no NTS is available at the end of the section.

In this section, we consider as an illustration a nuisance signal $\bd$ in Eq. \eqref{hyp} that encapsulates a magnetic signature that is modulated with the stellar rotation period plus long-term trends that are modulated with the stellar cycle. 
 Numerous techniques have been developed to estimate the contribution of magnetic activity to RV data (e.g., see \citeads{2021MNRAS.503.1248A}). We assume that an estimate of the noise activity signature has been obtained by one of these methods in the form of an ancillary time series for $\bd$.
{This ancillary time series is assumed to be the $\log{\rm R'_{HK}}$ activity indicator in the following (but other indicators can be used as well, see Table~\ref{tabex}).} \\
The noise component $\bn$ in Eq. \eqref{hyp} is composed of correlated granulation noise plus WGN. An NTS of $\bn$ composed with $L$ simulated RV time series obtained from MHD simulations is assumed to be available, except in the last part of this section, in which we relax this assumption.

The remaining section contains four parts. The first two parts describe the synthetic dataset with the computation of oracle $p$-values, and the validation of the $p$-value estimation algorithm with the impact of the priors on the computed $p$-values. In the third part, we study improvements that can be achieved by using generalized extreme values (GEV) to speed up the Monte Carlo simulations in algorithm~\ref{fig_algo3}. In the last part, we consider the particular (but practical) case in which no NTS is available for $\bn$ in Eq. \eqref{hyp}.
 
\subsection{{Synthetic RV time series and oracle $p$-values}}
\label{sec52}

{We first created synthetic RV time series used to i) generate an RV test time series that was later used in the input of algorithm~\ref{fig_algo1}, and ii) compute the oracle $p$-values that were later used to evaluate the accuracy of the $p$-value estimates derived from algorithm~\ref{fig_algo3}. These time series entail magnetic, granulation, and instrumental noises generated as described below}.

{Component $\bd$ was obtained as the realization of a Gaussian process (GP; \citeads{Rasmussen:2005:GPM:1162254}) of kernel 
\begin{equation}
k(t_i,t_j) =  \alpha_{GP} ~k_E(t_i,t_j) + k_M(t_i,t_j), 
\label{eq_kernel}
\end{equation}
with $k_E$ and exponential sine square kernel defined as  
\begin{equation}
k_E(t_i,t_j) = \exp\Bigg\{-\Gamma_{GP} \sin^2\Big(\pi |t_i-t_j| P_{GP}^{-1}\Big)\Bigg\},
\label{eq_GP1}
\end{equation}
and $k_M$ a Matérn-$3/2$ kernel defined as
\begin{equation}
k_M(t_i,t_j) = \Big(1+\sqrt{3}~|t_i-t_j|~\lambda_{GP}^{-1}\Big) ~\exp\Big\{- \sqrt{3}~|t_i-t_j|~\lambda_{GP}^{-1}\Big\},
\label{eq_GP2}
\end{equation}
with $\{t_i, t_j\}$ the time coordinates at indexes $i,j=1,\hdots,N$.
The hyperparameters of kernel $k$ in Eq. \eqref{eq_kernel} were set to $\alpha_{GP}=2$, $\Gamma_{GP}=22$ is the scale parameter, $P_{GP}=25$ days is the rotation period, and $\lambda_{GP}=1.609$ is the characteristic length scale. 
This GP was implemented using the \texttt{GEORGE} Python package\footnote{\url{https://github.com/dfm/george}} (see \citetads{2015ITPAM..38..252A} and online documentation\footnote{\url{https://george.readthedocs.io/en/latest/user/kernels/}} to find Eqs. \eqref{eq_GP1} and \eqref{eq_GP2}). Similar equations for kernels \eqref{eq_GP1} and \eqref{eq_GP2} can also be found in Eq.(8) of  \citetads{2019MNRAS.490.2262E} and Eq.(4) of \citetads{2022MNRAS.509..866B}, for instance.}

This GP model with the same parameters was also used to compute the synthetic ancillary time series $\vect{c}:= [c(t_1),\hdots,c(t_N)]^\top$, to which we added a WGN. 

For component $\bn$, we created a stochastic colored noise time series with a Harvey function and a WGN. The power spectral density of this noise model has the form (\citeads{1985ESASP.235..199H}, \citeads{2015A&A...583A.118M}) 
\begin{equation}
S_H(\nu_k^{+}; \vect{\theta}_\bn) := \Bigg( \frac{\frac{2\sqrt{2}}{\pi}  \frac{a_H^2}{b_H}}{1+\Big(\frac{\nu_k^{+}}{b_H}\Big)^{c_H}}\Bigg) + \sigma_\bw^2,
\label{eq_harvey}
\end{equation}
where the vector $\vect{\theta}_H = [a_H, b_H, c_H, \sigma_\bw^2]^\top$ collects the  amplitude, characteristic frequency, and power of the Harvey function, and  $\sigma_\bw^2$  is the variance of the WGN. Notation $\nu_k^{+}$ means that positive frequencies are considered.  
We calibrated the parameter vector $\vect{\theta}_H$ using a fit of the available MHD RV time series (see Sec.~\ref{sec4}). In this case and in the following, we used the \texttt{lmfit} Python package \citep{lmfit}, which makes nonlinear optimizations for curve-fitting problems. This choice was made for simplicity and because the algorithm is fast. For practical RV analyses, the user can optimize the fitting procedure to use a proper Markov chain Monte Carlo estimation procedure, for example. 
 
\begin{figure}[t!] \centering
\resizebox{0.75\hsize}{!}{\includegraphics{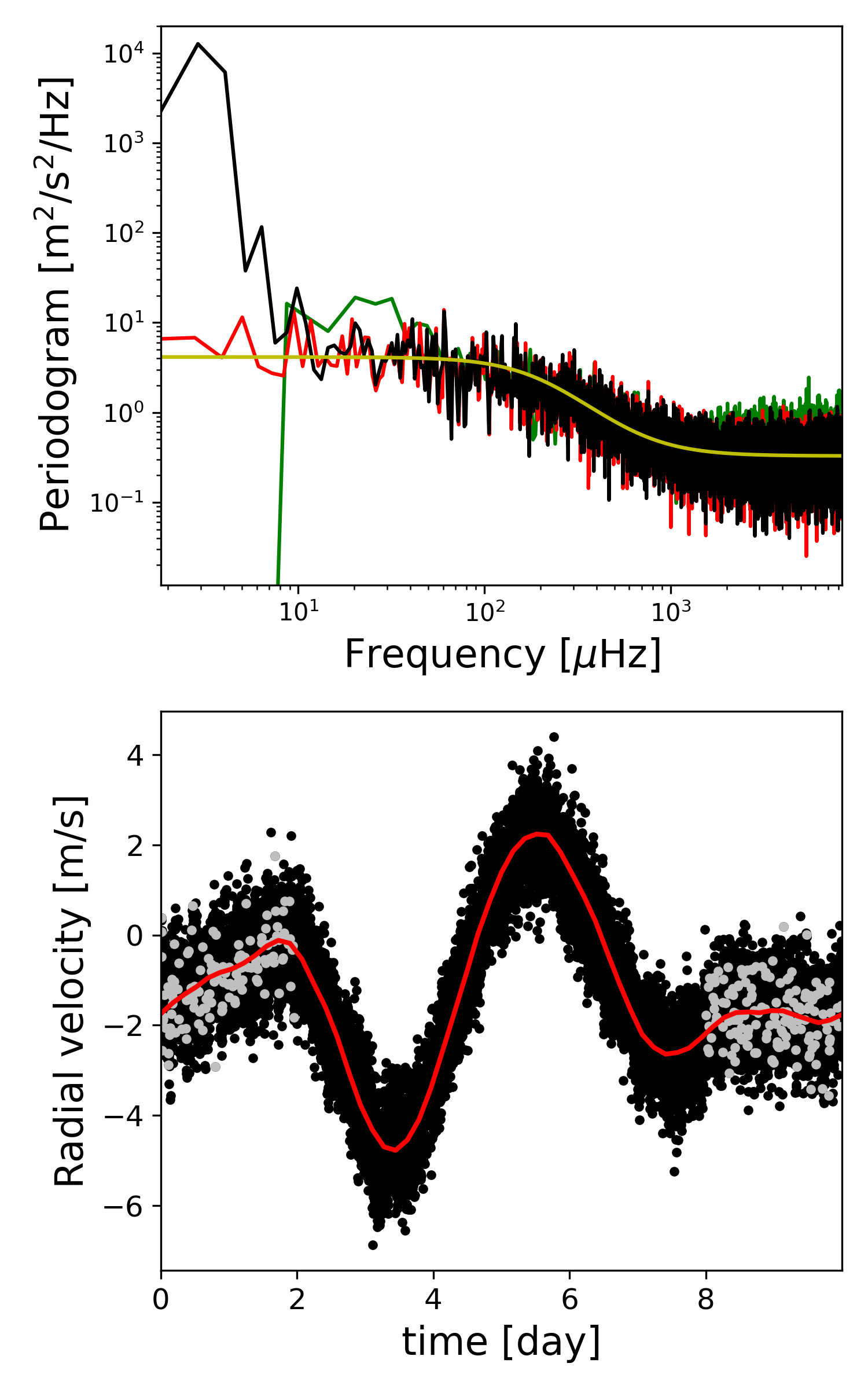}}
\caption{Example of the synthetic time series generated to validate the $\textsf{3SD}$ approach when noise equals magnetic activity plus long-term trends, granulation, and WGN. \textbf{Top}: Averaged periodograms of the $\text{two}$-day GOLF solar observations used in Sec.~\ref{sec23} (green) and the associated MHD training data (red), Harvey model fit to the MHD PSD (yellow), and averaged periodogram of the final synthetic data (GP+Harvey, black). \textbf{Bottom}: Example of synthetic RV data generated with a regular sampling (black) based on a GP noise modeling for the magnetic activity (red). For the study, the dataset was also sampled irregularly ({gray}).
}
\label{fig8}
\end{figure}

Synthetic time series with a PSD that is exactly given  by Eq. \eqref{eq_harvey} can be simulated with the Fourier transform (FT). To compute a synthetic time series from Eq. \eqref{eq_harvey},  we generated a WGN ${\bf w} \sim{\cal{N}}({\bf0},\boldsymbol{I})$, computed the fast FT of ${\bf w}$, performed a point-wise multiplication with $\sqrt{{\bf S}_H(\nu_k),}$ and finally took the real part of the inverse FT. An example of a synthetic $\text{ten}$-day-long time series obtained using models \eqref{eq_kernel} and \eqref{eq_harvey} is shown in Fig.~\ref{fig8} (black). From this time series, we created a large central gap between day $2$ and day $8$ in the time-sampling grid, and we randomly selected $N=288$ data points in the two remaining intervals (gray). This synthetic RV time series is the series that we test in the next three subsections.

To compute the oracle procedure (which has knowledge of the exact parameters that were used to generate the synthetic time series, i.e., $\vect{\theta}_H$ in Eq. \eqref{eq_harvey} and $[\alpha_{GP},\Gamma_{GP}, P_{GP}, \lambda_{GP}]$ in Eq. \eqref{eq_kernel}), we generated $10^4$ synthetic noise time series as described above, and we ran algorithm~\ref{fig_algo1} for each of the series.
The resulting oracle ground truth is shown in green in Fig.~\ref{fig5} (all panels), Fig.~\ref{fig_priors}, and Fig.~\ref{fig5b}.

\subsection{{Validation of algorithm 2}}
\label{sec53}

\begin{figure*}[ht!]
\resizebox{\hsize}{!}{\includegraphics{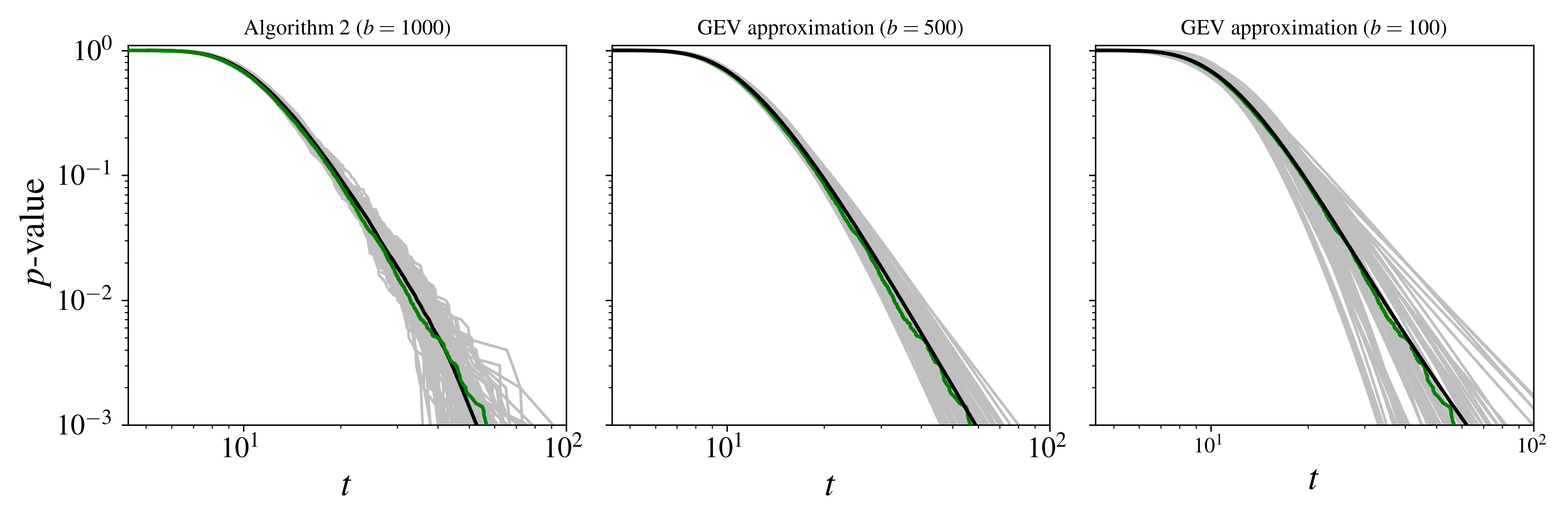}}
\caption{
{
$p$-value estimates resulting from algorithm~\ref{fig_algo1} when noise equals magnetic activity plus long-term trends, granulation, and  WGN, and where an NTS is available for granulation plus WGN. The mean $p$-value is shown in black. The oracle, which is the same for all three panels, is shown in green. From left to right: algorithm~\ref{fig_algo3} computed without the GEV, and with the GEV approximation with $b=500$ and $b=100$.}
}
\label{fig5}
\end{figure*}

We applied the \textsf{3SD} procedure on one of the synthetic time series. For this purpose, we assumed a realistic situation in which the noise models ${\cal{M}}_\bd$ and ${\cal{M}}_\bn$ that are used as inputs of algorithms~\ref{fig_algo1} and \ref{fig_algo3} are different from the models that are used to generate the synthetic RV dataset in Sec.~\ref{sec52}. 

Following an example given in \citetads{2012Natur.491..207D} and \citetads{2021MNRAS.503.1248A}, we used for $\calM_\bd$ a simple parametric model\footnote{{This model is currently implemented in the code (see Appendix.~\ref{appA}). Other models for ${\cal{M}}_d$, such as Gaussian process, are not implemented yet mainly for reasons of computational cost.}} 
composed of a linear combination of $\vect{c}$ and three low-order polynomials, 
\begin{equation}
\bd(t_j\;|\;\btheta_\bd) =  \beta_d  \cdot  {\bf{c}}(t_j)  + \epsilon_d \cdot t_j^2 + \zeta_d \,\cdot\;  t_j + \eta_d,~~~~~~j=1,\hdots,N
\label{model_a}
,\end{equation}
with $\vect{\theta}_\bd=[\beta_d, \epsilon_d, \zeta_d, \eta_d]^\top$ the corresponding parameter vector.

{For the available NTS ${\cal{T}}_L$, we selected the $L=5$ MHD time series described in Sec.~\ref{sec421}.
Using the LSP for \textsf{P}, the Max test for \textsf{T}, ${\cal{T}}_L$, the model \eqref{model_a} for $\calM_\bd$, no model $\calM_\bn$, and the synthetic ancillary series  $\vect{c}$, we then computed algorithm~\ref{fig_algo1} and obtained a test value $t$. 

To compute algorithm~\ref{fig_algo3}, the inputs parameters (\textsf{P}, \textsf{T}), $\calM_\bd$, and $\vect{c}$ were the same as for algorithm~\ref{fig_algo1}. 
For the additional inputs, we took the result from row 2 of algorithm~\ref{fig_algo1} to obtain $\widehat{\vect{\theta}}_\bd $, the vector of (four) parameters estimates, along with its scale parameter $\widehat{\bdelta_\bd}$. We selected a uniform prior $\pi_\bd$ for the noise parameters with interval widths set to twice the estimated standard error.
We selected an AR model \eqref{eq_AR} for model ${\cal{M}}_\bn$ (as in Sec.~\ref{sec41}), and we fit this model to the NTS to obtain the estimated parameters $\widehat{\vect{\theta}}_\bn$. 
We finally set up the size of the Monte Carlo simulations to $B=100$ and $b=1000$.
The $p$-value estimates resulting from algorithm~\ref{fig_algo3} are shown in the left panel of Fig.~\ref{fig5}. When they are compared with the oracle $p$-values, algorithm~\ref{fig_algo3} is shown to provide reliable $p$-value estimates at any $t$ for colored noise and activity.}\\

\begin{figure}[t]
\resizebox{\hsize}{!}{\includegraphics{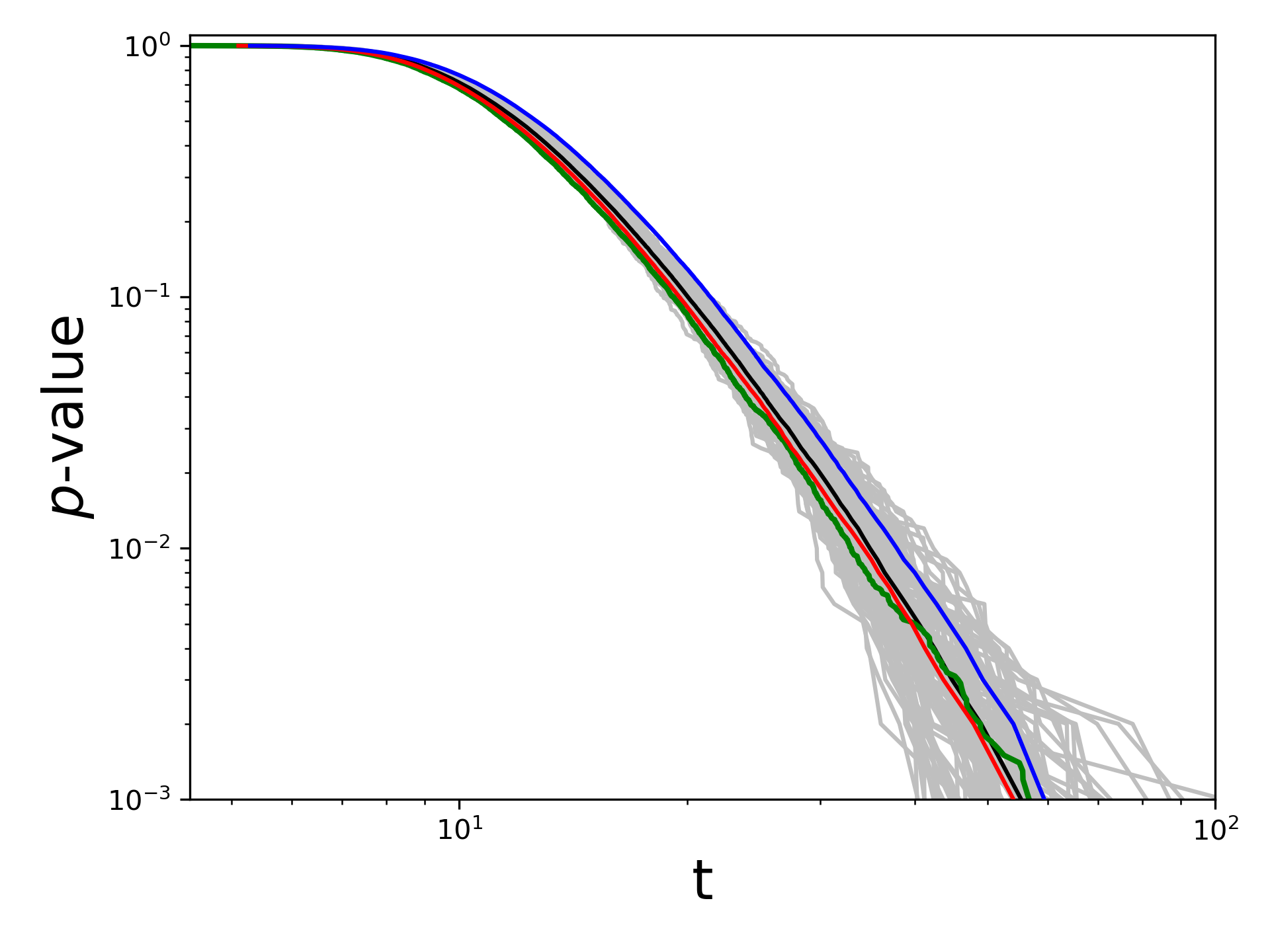}}
\caption{
{$p$-value estimates resulting for different choices of the priors $\pi_\bd$ in row 13 of algorithm~\ref{fig_algo3}. Gray and black show the results for a  Gaussian prior centered on the estimated parameter values with the scale parameters taken as $3\;\sigma$, with $\sigma$ the estimated standard deviation of the parameter estimates. Gray shows examples of $p$-values $\widehat{p}_v(t)$, and black shows the mean $p$-value. Red and blue show results for uniform priors with intervals centered on the estimated parameter values and widths taken as $1\;\sigma$ (red) and  $10\;\sigma$ (blue). The true (oracle) $p$-value is shown in green.}
}
\label{fig_priors}
\end{figure}

{As a final validation step of algorithm~\ref{fig_algo3}, we finally examined the impact of the priors $\pi_{\bf{d}}$ (involved in row 13) on the derived $p$-values. The results are shown in Fig.~\ref{fig_priors}. 
This example shows a weak impact, with an accurate estimation for any $t$ in almost all cases. At the extreme of $10~\sigma$ for the perturbation interval width, the $p$-value estimate becomes more conservative: at a given value of $t,$ the estimated $p$-value is biased upward (blue curve).  In practice, the influence of the  chosen priors on the estimated $p$-values can always be confirmed by a  simulation study of this type.}

\subsection{Generalized extreme values}
\label{sec54}

Algorithm~\ref{fig_algo3} does not rely on any model for the estimated CDF {$\widehat{\Phi}_{T|{\cal{H}}_0}$} of the test statistics in Eq. \eqref{T_Max}. 
For the estimated mean $p$-value to be accurate, it therefore requires the parameter $b$ involved in algorithm~\ref{fig_algo3} to be large (typically $b\geq 1000$ or more). As $B\times b$ Monte Carlo simulations are involved (with $B$ typically $\geq 100$), algorithm~\ref{fig_algo3} is  computationally expensive.
Fortunately, univariate extreme-value theory shows that the maximum of a set of identically distributed random variables follows a GEV distribution \citep{Coles_2001}. This suggests that GEV distributions can be used as a parametric model for the CDF of Eq. \eqref{T_Max} to improve the estimation accuracy.
This method was used by \citet{2014MNRAS.440.2099S} when only white noise is involved in model \eqref{hyp} (i.e., no colored noise or nuisance signal $\bd$). \citet{2017arXiv170606657S} extended the work of \citet{2014MNRAS.440.2099S} to the case of colored noise. Here, we evaluate the gains that this parameterization  can contribute to the bootstrap procedure for the general case of colored noise and magnetic activity signals.

The GEV CDF depends on three parameters: the location $\mu \in \mathbb{R}$, the scale $\sigma \in \mathbb{R}^{+}$ , and the shape $\xi \in \mathbb{R}$,
\begin{equation}
        \label{eq_GEV}
        G(t) = \mathrm{e}^{-\Big[ 1+\xi \Big( \frac{{t}-\mu}{\sigma}\Big)\Big]_+^{-\frac{1}{\xi}}},
\end{equation}
with $[x]_+ := \max(0,x)$. When $\xi \neq 0$, $G(t)$ in Eq. \eqref{eq_GEV} is only defined for $t = 1+\xi \frac{t-\mu}{\sigma} >0$. When $\xi=0$, $G(t)$ is derived by taking the limit at $t=0$.

In algorithm~\ref{fig_algo3}, the unknown parameters $\btheta_{GEV} := [\xi, \mu, \sigma]^\top$ can be estimated using the $j=1,\hdots,b$ test statistics $t(\bx^{\{i,j\}})$, for instance, by maximum likelihood. In this case, the maximization can be obtained by an iterative method \citep{Coles_2001}.
This leads to $i=1,\hdots,B$ GEV parameters $\widehat{\btheta}_{GEV}^{\{i\}} := [\widehat{\xi}^{\{i\}}, \widehat{\mu}^{\{i\}},\widehat{\sigma}^{\{i\}}]^\top$, from which the $p$-value can be estimated by ${\widehat{p}_{v}^{\{i\}}(t)}:=1-G(t | \widehat{\btheta}_{GEV}^{\{i\}})$.

{The results of the GEV parameterization are shown in the middle and right panels of Fig.~\ref{fig5} (for $b=100$ and $b=500$)}. The $p$-value estimates resulting from the GEV approximation are still given in a fairly tight $90\%$ interval around the oracle. Moreover, when only half of the Monte Carlo realizations are used compared to the initial $b$ value involved in algorithm~\ref{fig_algo3} ($b=500$ instead of $1000$), a similar precision of the $p$-value estimates is obtained. {Hence, {the GEV approximation allows us to halve the computational time of algorithm~\ref{fig_algo3} for the same precision} (case $b=500$). {Reducing} the computational time by ten ($b=100$) leads to larger $90\%$ intervals at constant $p$-value (e.g., the interval is $\sim 40$\% larger at a {$p$-value} of $1$\% in the right panel of Fig.~\ref{fig5}).}

\subsection{{Algorithm 2 without NTS (practical case)}}
\label{sec55}

{We consider now the common situation in which no NTS is available. In this case, the parameters of  noise $\bn$ can be estimated from the data (maybe along with the magnetic activity signal $\bd$) through a model $\calM_\bn$. Standardization is then performed by generating synthetic time series according to $\calM_\bn$ in algorithm~\ref{fig_algo1}.

When no NTS is available, two steps in algorithm~\ref{fig_algo3} deserve particular attention. First, in rows 3 and 4, we set $L=1$ because the noise parameters estimate $\widehat{\btheta}_\bn$ of model ${\cal{M}}_\bn$ is based on the unique time series under test $\bx$. Next, in row 7,  as many synthetic time series can be generated that follow model $ \calM_\bn(\btheta_\bn)$ as necessary. Hence, in principle, $L\rightarrow\infty$. In practice, however, the choice of $L$ is a compromise between estimation error and computation time. For the purpose of making massive Monte Carlo simulations, we used the low value $L=5$.

 As in Sec.~\ref{sec53}, we applied the \textsf{3SD} procedure to one of the synthetic time series described in Sec.~\ref{sec52}. We first ran algorithm~\ref{fig_algo1} with model ${\cal{M}}_\bd$ such that
\begin{equation}
\bd(t_j\;|\;\btheta_\bd) =  \beta_d  \cdot  {\bf{c}}(t_j),~~~~~~j=1,\hdots,N.
\label{model_b}
\end{equation}
For model ${\cal{M}}_\bn$, a Harvey function \eqref{eq_harvey} was used. There is no NTS and the pair (\textsf{P} and \textsf{T}) was taken as the LSP periodogram and the Max test. 

To compute algorithm~\ref{fig_algo3}, the input parameters (\textsf{P}, \textsf{T}), $\calM_\bd$, $\calM_\bn$ , and $\vect{c}$ are the same as for algorithm~\ref{fig_algo1}. 
For the additional inputs, we obtain $\widehat{\vect{\theta}}_\bd $ and $\widehat{\bdelta_\bd}$ from row 2 of algorithm~\ref{fig_algo1} and set $\pi_\bd$ as in Sec.~\ref{sec53}.
Parameters $\widehat{\vect{\theta}}_\bn $ were obtained from row 9 of  algorithm~\ref{fig_algo1} based on a fit of the LSP of $\bx$ (as in \citetads{2011A&A...525A.140D}; this implicitly assumes that the LSP provides an acceptable approximation of the true PSD). 
We finally set up the size of the Monte Carlo simulations as $B=100$ and $b=1000$.
The $p$-value estimates resulting from algorithm~\ref{fig_algo3} are shown in Fig.~\ref{fig5b}. The estimated mean $p$-values and true $p$-values match very well, similarly to the case with an NTS (Sec.~\ref{sec53}, see left panel of Fig.~\ref{fig5}). This shows that the $p$-value can be robustly estimated even when no NTS is available.

{
We anticipate, however, that the performances of detection procedures based on estimates of the covariance structure from the data will probably depend on the RV data (noise sources, data sampling, etc.). The comparison of specific cases is left to a future paper.
}

\begin{figure}[t!]
\resizebox{\hsize}{!}{\includegraphics{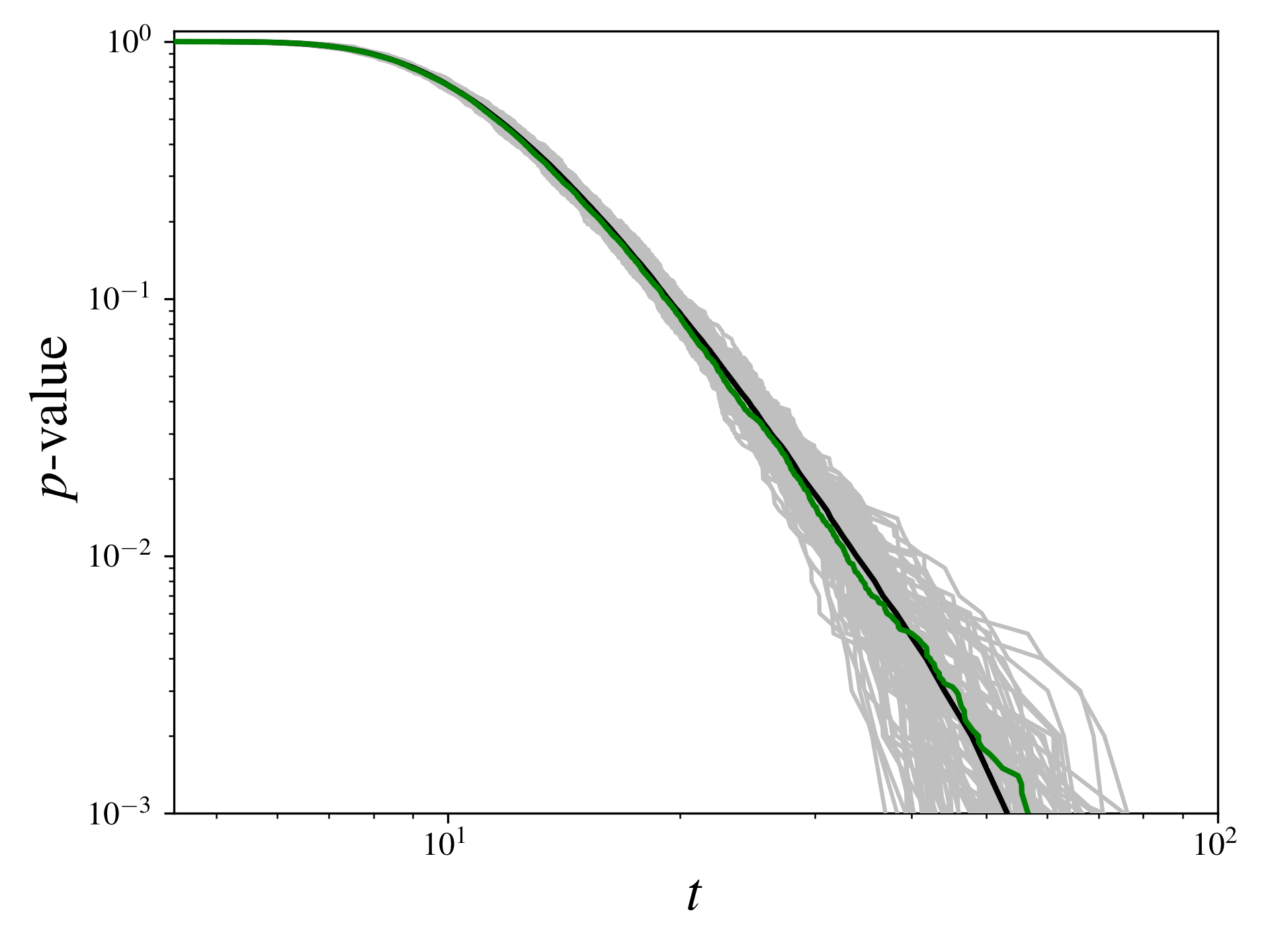}}
\caption{
{Same as Fig.~\ref{fig5}, but for a practical case without NTS for $\bn$ in Eq. \eqref{hyp}. In this case, the covariance structure of the stochastic noise source is estimated from the data with model ${\cal{M}}_\bn$ given in Eq. \eqref{eq_harvey}.}
}
\label{fig5b}
\end{figure}

\section{Application to RV exoplanet detection: The debated detection of $\alpha$CenBb}
\label{sec6}

Alpha Centauri Bb is a debated $\sim1.13$ Earth-mass  (minimum mass) planet whose detection was reported around  our solar neighborhood  star $\alpha$CenB  with an orbital period of $3.2$ days \citepads{2012Natur.491..207D}. It is challenging to achieve the detection of such a small planet because $\alpha$Cen  is a triple star and the planetary RV semi-amplitude ($\sim0.51$ m/s) was at the limit of the instrumental precision of HARPS ($\sim1$ m/s). 
 The observed RV campaign contains $N=459$ points, irregularly sampled during a time span of $\text{four}$ years between 2008 and 2011. The dataset contains large gaps between the observation years as a result of the visibility of the star, and small gaps that are due to the available nights.
 
{The planet detection has been debated in a lively way: using  analysis tools different from the detection paper in \citetads{2013ApJ...770..133H}, using different stellar noise modeling in \citetads{2016MNRAS.456L...6R}, and recently using  a different technique of randomization inference to measure the $p$-value of the periodogram peak at $3.2$ days by \citetads{2021arXiv210514222T}.}

In this section, we analyze this RV dataset as a case study for the \textsf{3SD} procedure.  {The objective of the second part (Sec.~\ref{sec62}) is also to demonstrate that algorithm~\ref{fig_algo3} offers the possibility of investigating the effects of model errors on the frequency analysis and on the estimated $p$-values. The purpose is not to argue for or against a chosen noise model, but to study the effects of model mismatch.}
We first present the results of the \textsf{3SD} procedure computed with the noise models involved in the detection paper \citepads{2012Natur.491..207D}. We then analyze the robustness of the results using a different  noise model.

\subsection{{\textsf{3SD} procedure with models from the detection paper}}
\label{sec61}

In short, the initial model ${\cal{M}}_\bd$ described in the detection paper \citepads{2012Natur.491..207D} contains three noise components. 
First, it contains the binary contribution (${\bf{m}}_B$) of $\alpha$Cen A, modeled with a second-order polynomial ($\text{three}$ free parameters). 
Second, it contains the magnetic cycle, modeled with the linear relation of Eq. \eqref{model_b} and ${\rm log R'_{HK}}$ (for which periods $<40$ days have been filtered out) taken as ancillary series  (${\bf{m}}_{C}$, $\text{one}$ free parameter). 
Third, it contains the magnetic (rotation modulated) activity (${\bf{m}}_M$), modeled as the sum of sine and cosine waves fitted at the stellar rotation period ($\sim 37$ days) and its first four harmonics.    Because of the stellar differential rotation, the authors fitted each of the four years of observations individually to allow the stellar rotation period estimate to vary slightly from year to year. They also used a different number of harmonics for each of the four RV subsets. We refer to the supplementary material' of \citetads{2012Natur.491..207D} for the global model that is fit to each of the four RV subsets (see their Eqs. p.7). In total, ${\bf{m}}_{M}$ contains $19$ free parameters.  

The sum of components ${\bf{m}}_B$, ${\bf{m}}_C$ , and ${\bf{m}}_M$ forms the noise component $\bd$ in model \eqref{hyp}. The authors assumed that component $\bn$ in \eqref{hyp} is uncorrelated and there is no NTS. 
 
\noindent We applied the \textsf{3SD} procedure to the RV data of $\alpha$CenB with the following setting for algorithm~\ref{fig_algo1}:
\begin{itemize}
    \item the Lomb-Scargle periodogram as \textsf{P},
    \item the Max test as \textsf{T},
    \item the set of linearly sampled frequencies as $\Omega:=[9.4\times 10^{-4},215.8]$ $\mu$Hz (with $\Delta \nu = 1.88\times 10^{-3}$ $\mu$Hz, and $N_\Omega=114750$ frequencies in total), 
    \item no training time series or parametric model for the stochastic noise component $\bn$ (${\cal{T}}_L=\emptyset$; ${\cal{M}}_\bn=\emptyset$), 
    \item the model defined in \citetads{2012Natur.491..207D} for component $\bd$, recalled above as ${\cal{M}}_\bd$, 
    \item the ancillary time series ${\bf c}$ as the $40$-days-filtered ${\rm log R'_{HK}}$ time series available from the data release of \citetads{2012Natur.491..207D}.
\end{itemize}
In row 11 of algorithm~\ref{fig_algo1}, we obtained a standard deviation $\widehat{\sigma}_\bw$ of the data residuals of $1.23$ m/s\footnote{For these data, different error bar estimates $\widehat{\sigma}_i$ are provided and were considered in the parameter estimation. For simplicity, algorithm~\ref{fig_algo3} used a common uniform prior with an interval equal to $\widehat{\sigma}_\bw$ , but independent priors matched to $\widehat{\sigma}_i$ can be implemented here.}, in agreement with the value given in \citetads{2012Natur.491..207D}. The standardized periodogram computed in row 14  is shown in Fig.~\ref{appli_fig1}. The algorithm computed a test value  $t=16.03$ (peak at $3.2$ days).

\begin{figure}[t!]
\resizebox{\hsize}{!}{\includegraphics{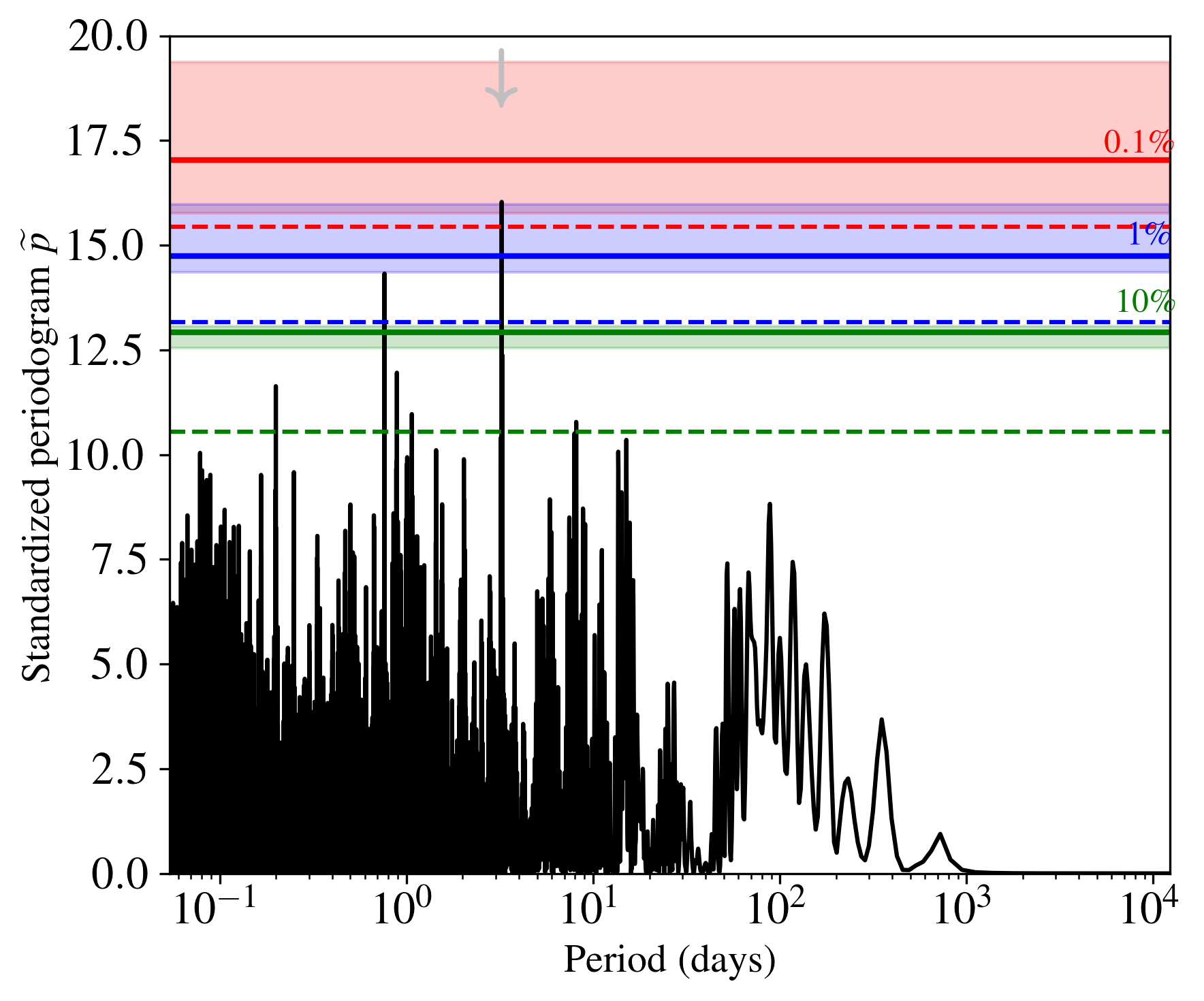}}
\caption{
{Standardized Lomb-Scargle periodogram of the $\alpha$ CenB RV residuals. The noise model ${\cal{M}}_{d}$ used to create these residuals is taken from \citetads{2012Natur.491..207D}. 
$p$-values estimated by a classical bootstrap procedure (see text) are shown with the dashed lines. Mean $p$-values derived with algorithm~\ref{fig_algo3} are shown with solid lines, and their credibility interval with the shaded regions. The peak at $3.2$ days is shown by the gray vertical arrow.} 
}
\label{appli_fig1}
\end{figure}

We then ran algorithm~\ref{fig_algo3} to evaluate the $p$-value of $t$ with same inputs as algorithm~\ref{fig_algo1} that are listed above. \sop{In addition, we plugged into algorithm~\ref{fig_algo3}} a number of Monte Carlo simulations of $B=100$ and $b=1000$, a Gaussian prior for $\pi_\bw$ with a standard deviation of $0.04$ m/s ($1\sigma$ estimated error bar), and  a uniform prior $\pi_\bd$ with a $1\sigma$ interval width on each of the $23$ free parameters of model ${\cal{M}}_\bd$. 

The $t$ values corresponding to mean $p$-values of $10$, $1,$ and $0.1$\% are shown with solid green, blue, and red curves, respectively, in Fig.~\ref{appli_fig1}. The $90\%$ intervals are shown with the corresponding colored shaded regions. 
{The value $t:=\widetilde{{\bf{p}}}(3.2\;d)= 16.03$ corresponds to an estimated mean $p$-value of $0.47$\% with a $90$\% interval of $[0.1\%, 0.7\%]$}.
We confirmed that these results are reasonably stable with respect to the prior model $\pi_\bd$ for the ${\cal{M}}_\bd$ model parameters in row 13 of algorithm~\ref{fig_algo3} (e.g., a Gaussian instead of a uniform prior with interval widths set to $1\sigma$ standard error leads {to $\overline{\widehat{p}}_{v}(16.03) = 0.43$\% with $[0.1\%, 0.7\%] \;90\%$ interval)}.

 \citetads{2012Natur.491..207D} assumed that the data residuals (without a planet) are white. While the $p$-value evaluation procedure used in this paper was not specified, we assume that they used the classical bootstrap technique in which the $p$-values are evaluated by shuffling the dataset and keeping the observing time dates constant. The $t$ values corresponding to the $p$-values of $10$, $1,$ and $0.1$\% computed with this classical method are shown by dashed lines for comparison. The largest periodogram peak at $3.2$ days lies in the region of $p$-values below $0.1$\%, which is consistent with the result of the publication paper (a $p$-value of $0.02$\%).


We conclude from this analysis that if the model assumed by \citetads{2012Natur.491..207D} is accurate, the $3.2$ -days peak is associated with a mean $p$-value of $0.47$\%,  which is small but nevertheless about $23$ times larger than the $0.02\%$ $p$-value estimate given in the detection paper. In general, the $p$-value derived without taking the errors on the noise model parameters into account is underestimated.

\subsection{Systematic study of noise model errors}
\label{sec62}

\begin{figure*}[ht!]
\resizebox{\hsize}{!}{\includegraphics{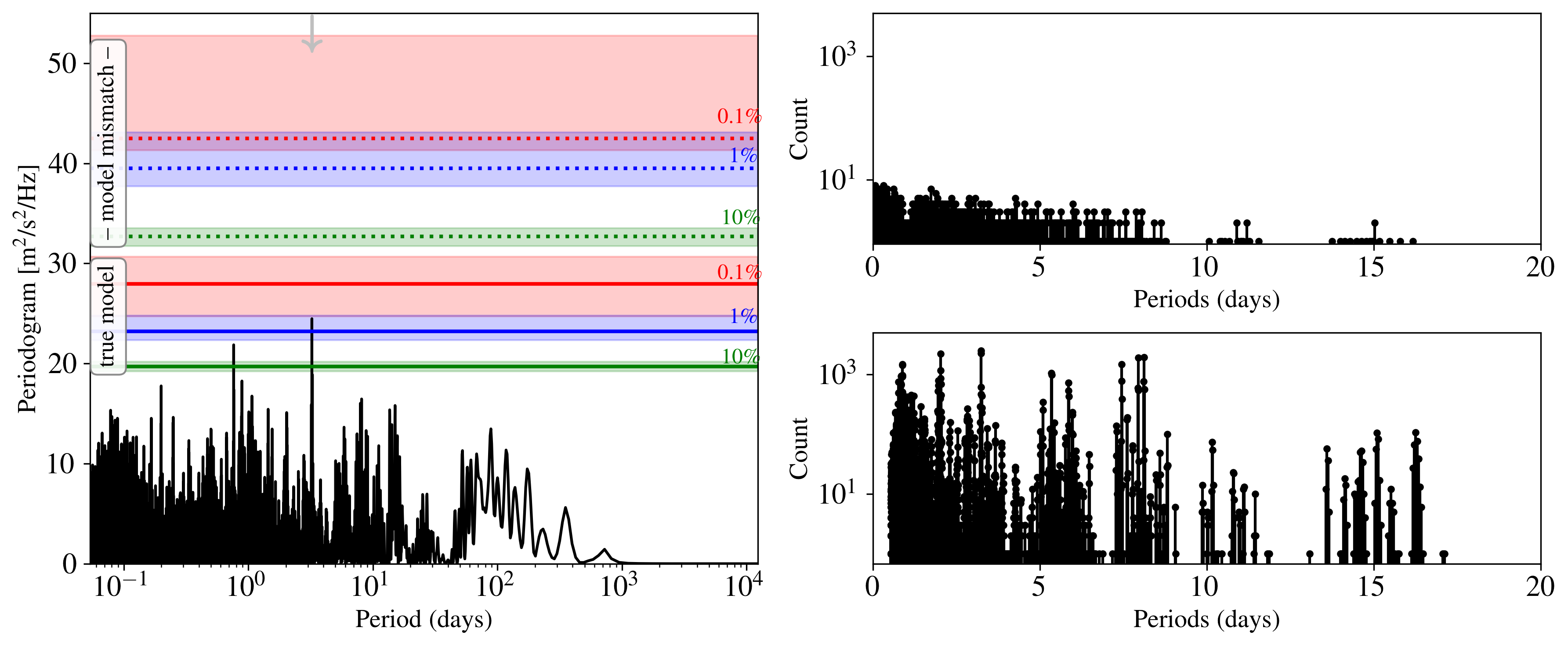}}
\caption{
{ \sop{Impact of noise model errors on the significance levels.} 
Left: Lomb-Scargle periodogram of the $\alpha$ CenB RV residuals. The peak at $3.2$ days is shown by the gray vertical arrow. 
Solid lines are the mean $p$-values estimated without a model error. The $p$-value at $t:={\bf{p}}(3.2\;d)= 24.48$ is within the $90\%$ interval $[0.1\%, 0.7\%]$.
Dotted lines are the mean $p$-values estimated with a model error (see text). The $p$-value at $t= 24.48$  this time is within the $90\%$ interval $[59.5\%, 64.7\%]$.
Right:  Distribution of the period index where the maximum periodogram value was found in row 15 of algorithm~\ref{fig_algo3} (zoom at periods $<$ 20 days). The standard version of the \textsf{3SD} procedure (simulating the situation without a model error) is shown in the top panel, and the hybrid version (detection process run with a model error) is shown in the bottom panel. A bias toward specific periods is visible in the bottom panel: the model error leaves residuals in which periodicities are imprinted.
}
}
\label{appli_fig2}
\end{figure*}

We considered another noise model ${\cal{M}}_\bd$ for $\bd$. Following \citetads{2016MNRAS.456L...6R} {(see their Sec. 4)}, we still considered a nuisance signal of the form $\bd = {\bf{m}}_B + {\bf{m}}_{C} +  {\bf{m}}_{M}$, but with some modifications.
 The estimation of the noise model parameters on the RV dataset was now performed in two steps. The binary and long-term trend components ${\bf{m}}_{B}$ and ${\bf{m}}_{C}$ were fit first. The magnetic activity component ${\bf{m}}_{M}$ was fitted in a second step on the data residuals as a Gaussian process with a quasi-periodic covariance kernel of the form of Eq. \eqref{eq_kernel}, with $k_M$ fixed to $0$ and parameters $[\alpha_{GP},\Gamma_{GP}, P_{GP}]$ free with $P_{GP}$ in  intervals of $30 \text{ and }42$ days. 

When we ran our detection algorithm~\ref{fig_algo1} with this new noise model, we obtained final data residuals of $0.78$ m/s standard deviation, and the peak at $3.2$ days disappeared. Similarly to \citetads{2016MNRAS.456L...6R}, we performed an injection test with a planetary signal at $3.2$ days with an amplitude of $0.51$ m/s to verify that the GP noise model does not overfit the planet signal. We found the periodogram peak of the data residuals to be the highest peak, confirming the observations of \citetads{2016MNRAS.456L...6R} that the detection outcome can be very sensitive to the choice of the noise model plugged in the detection algorithm.

Model errors are unavoidable in practice. For the same dataset, different experts may assume different noise models, and the magnitude of the absolute error resulting from these models can probably be a lower bound of the relative difference between the models.
These observations suggest that automatic studies of the robustness of the detection process to model errors should be incorporated in the process of estimating $p$-values. For this purpose, a hybrid version of the \textsf{3SD} procedure can be used.

This hybrid version simply consists {of two modifications of the \textsf{3SD} procedure. First, we allow in algorithm~\ref{fig_algo3}} that the procedure generates the noise-training series in rows 12 to 15 with a given model ${\cal{M}}_\bd$ (the Gaussian process model), and to fit it in row 16 with another model, for instance, ${\cal{M}}_\bd'$ (the model of \citetads{2012Natur.491..207D}).
{Second, we do not use the standardization step in algorithm~\ref{fig_algo1} (rows 5 to 13) because data residuals resulting from different models will have different dispersions, with a larger dispersion in presence of model error. Keeping standardization (i.e., division by the estimated variance) would then lead to relatively smaller ${{\bf{\widetilde{p}}\;|\;{\bf{\overline{p}}}}}_L$ components for the model with the highest residuals and artificially hide the effects of the mismatch we seek to detect and show.}  
The test values corresponding to $p$-values of $10$, $1,$ and $0.1\%$ computed in this hybrid configuration are shown with horizontal dotted lines in Fig.~\ref{appli_fig2} (left panel). {For comparison, the levels corresponding to the no model error configuration are shown with solid lines. The $90\%$ intervals are again shown with shaded regions.}
{We find that if the detection process indeed used the right noise model, the largest data peak at $3.2$ days has a $p$-value  of  $0.62\%$. If, however, the detection process committed a slight mismatch in the model error, the actual $p$-value  of the same peak increases to $62.02$\%. This means that peaks as large as the largest data peaks occur quite often, even without a planet, when the detection process is run under a model mismatch such as the one simulated here. Model mismatch is a likely situation in practice. A robust detection would be a detection whose $p$-value remains low under realistic model mismatches.}

{ Interestingly, the hybrid version of the \textsf{3SD} procedure further allows an investigation of the distribution of the period index where the $B\times b$ maximum values of $\bf{p}$ are found (see in Sec.~\ref{sec4}). This allows indicating periodicities that are not caused by planets and may be used as a warning in the detection process when it finds a planet at these frequencies.
This distribution is shown in Fig.~\ref{appli_fig2} (bottom right panel). 
In this case, the distribution of the frequency index of the largest periodogram peaks significantly moves away from the uniform distribution that is seen when the noise model is correct (top right panel). When the noise model is correct, each frequency has been randomly hit a few times  (the holes are caused by frequencies that are removed in the noise model). In contrast, when the noise model is not correct, the maximum peak was found $2504$ times at the $3.2$ -day period (compared with $\text{two}$ times under no model error), or $1917$ times at the $7.94$ -day period (compared with $\text{one}$ time without a model error).
}

In conclusion, when 1) the estimation errors on the noise  parameters {(assuming no model error)} are taken into account and 2) the effects of possible  realistic errors on the noise model are investigated, the degree of surprise brought by the peak at $3.2$ days decreases substantially. This study of  the $\alpha$ CenB dataset  supports the conclusion that was reached in previous works (\citeads{2013ApJ...770..133H}, \citeads{2016MNRAS.456L...6R}, \citeads{2021arXiv210514222T}), namely a likely false detection at $3.2$ days,{ but with a different analysis. Beyond the particular case of the 
$\alpha$ CenB dataset, accounting for possible noise model errors} appears to be a critical point, particularly when the amplitude of the planet signal is small compared to the nuisance signals.

\section{Discussion and conclusions}
\label{ccl}

 {Assessing the significance level of detection tests is critical for an exoplanet RV detection in the regime of a low signal-on-noise ratio and in presence of colored noise of unknown statistics.  Impacts of errors on the parameter estimation and of model errors on the $p$-values  have been little studied in the exoplanet community so far.}
 
 { To overcome this limitation, we have presented a new method for computing $p$-values of detection tests for RV planet detection in presence of unknown colored noise. The developed Bayesian procedure is based on the principles of statistical standardization and allows using some training samples (NTS), if available (Sec.~\ref{sec2}). It involves Monte Carlo simulations (Sec.~\ref{sec3}) and allows evaluating the robustness of the derived significance levels against specific model errors (Sec.~\ref{sec62}).  }
{Our procedure builds on and extends our previous study ( \citetads{2020A&A...635A.146S}), which was limited to the case of  regular sampling  and was not able to handle magnetic activity signals. This procedure} allows deriving in algorithm~\ref{fig_algo3} the $p$-value  corresponding to a test statistic produced in detection algorithm~\ref{fig_algo1} while accounting for uncertainties in the noise model parameters.

Periodogram standardization can be implemented using a (possibly simulated) NTS. {Examples of simulated NTS can come from MHD simulations of granulation, supergranulation, or any other stellar simulations that are shown to be reliable}. This opens future collaborations between stellar modelers and RV exoplanet scientists  (see Sec.~\ref{sec4}). The NTS can also come from instrumental measurements, independent of the studied RV dataset. 
In addition, the procedure can take advantage of ancillary data to deal with nuisance signals such as the stellar magnetic activity (e.g., ${\rm log R'_{HK}}$ time series).

Overall, the proposed detection procedure is versatile in i) the periodograms (or other frequency analysis tools), ii) the detection tests (classical or adaptive), and iii) the noise models that can be plugged in.
Examples of input parameters for the three items above are listed in Table~\ref{tabex}. {Additionally, Python codes are released on GitHub (see Appendix.~\ref{appA})}.
Although not emphasized in this paper, the \textsf{3SD} procedure also opens the possibility of using adaptive tests such as the higher criticsm or the Berk-Jones tests (\citeads{2004math.....10072D}, \citeads{Berk_1979}). These tests are more powerful than classical detection tests (as the Max test presented in Eq. \eqref{T_Max}) in the case of exotic  planet signatures of small amplitudes (e.g., planets of high eccentricity or multiplanet systems) when the data sampling is regular (see \citeads{2017ITSP...65.2136S}). How these adaptive tests may be incorporated in the proposed \textsf{3SD} procedure to evaluate their performance in the context of irregular time sampling will be investigated in a further study.

We have illustrated how the procedure can be implemented in the presence of  instrumental, granulation, and magnetic activity, of long-term drifts of stellar origin (magnetic cycle), and binary noises in model \eqref{hyp} (Sec. ~\ref{sec4}, \ref{sec5}, and \ref{sec6}). Other RV noise sources can similarly be considered in the procedure when they are identified in the RV data. 

Implementing the \textsf{3SD} procedure may present some issues, but these can at least partially be solved.
 First, the procedure can be very time consuming owing to several factors (generation of NTS simulations, large number of Monte Carlo simulations, and the complexity scales with the number of parameters of the noise models). We have shown that the GEV framework allows us to reduce this computational cost (Sec.~\ref{sec54}).  
 Second, the computed $p$-values remain dependent on the choice of the input noise models. While this is not specific to our approach, we showed that a hybrid version of the procedure allows us to quantify the robustness of  $p$-value estimates to specific model changes (Sec.~\ref{sec62}).
Third, the computed $p$-values are more conservative than the $p$-values obtained by random shuffling of the best-fit residuals, as is often done in RV data analysis (Sec.~\ref{sec61}). This is because they take the main errors into account that are generated during the detection process (estimation error on the parameters and/or specific model error). 
Last, when the detection process makes some error in the assumed noise  model, estimating and removing a nuisance signal from the RV dataset may lead to a nonuniform distribution of the frequency index where the largest periodogram peak is found (this is visible in the bottom right panel of Fig.~\ref{appli_fig2}). The reason is that imperfect subtraction of nuisance signals (caused by a  model error) leads to residual long-period signals in which the periodicities of the sampling grid are imprinted.  Model errors always remain in practice, and the proposed algorithm allows diagnosing the effects of some of them.

Finally, throughout this paper, we have illustrated the proposed method on solar data, synthetic datasets, and HARPS data of $\alpha$CenB.  The next step is to analyze some still-to-be confirmed low-amplitude planet detection.


\begin{acknowledgements}

The authors would like to thank the referee for her/his helpful comments, which led to improved versions of this study. 
S. Sulis acknowledges support from CNES. D. Mary acknowledges support from the GDR ISIS through the \textit{Projet exploratoire TASTY}. 
MHD stellar computations have been done on the ``Mesocentre SIGAMM'' machine, hosted by \textit{Observatoire de la C\^ote d'Azur}.
  
The GOLF instrument onboard SOHO is a cooperative effort of scientists, engineers, and technicians, to whom we are indebted. SOHO is a project of international collaboration between ESA and NASA. 
We acknowledge X. Dumusque, Astronomy Department Geneva Univ., Switzerland, for the public HARPS-N data that have been made available at http://cdsarc.u-strasbg.fr/viz-bin/cat/J/A+A/648/A103.
We acknowledge support from PNP and PNPS-CNES for financial support through the project \textit{ACTIVITE}.

\end{acknowledgements}


\bibliographystyle{aa} 
\bibliography{Bibfile} 

 \begin{appendix} 

\section{{Algorithms available on GitHub}}
\label{appA}

{
We do not describe the algorithms here (see Sec.~\ref{sec24} and \ref{sec32}), but rather the contents of the directory that is available online\footnote{\url{https://github.com/ssulis/3SD}}. This directory contains algorithms  written in Python 3.

In the directory ``\textit{Func/}'', the user can find the implementation of algorithms~\ref{fig_algo1} and ~\ref{fig_algo3}. 
\sop{He/she can also find one function to evaluate the periodograms $\bp$ and $\overline{\bp}_L$ and one to evaluate the detection tests involved in algorithm~\ref{fig_algo1}.
We also implemented several functions in a file to model a set of nuisance signal models ${\cal{M}}_d$, estimate the parameters ${\boldsymbol{\theta_d}}$ of these models, and generate synthetic data from these models. 
Similarly, another file has been written for the set of nuisance signal models ${\cal{M}}_n$. 
Finally, the reader can find a file containing functions that save and read the results of the outputs of algorithm~\ref{fig_algo3}. }
        
The current release implements the different setting (e.g., periodograms, considered set of frequencies, detection tests, and estimation of the noise model parameters) that were used in this work. 
Further settings will be added in the future (see Table.~\ref{tabex}), but the users can already implement their own tools, such as new periodograms,  detection tests, nuisance component models, parameter estimation procedure,  prior laws, or stochastic component models.
\sop{Note that the files containing algorithm~\ref{fig_algo1},  algorithm~\ref{fig_algo3}, and the read/save functiondo not need to be modified by the user. }
         
As an illustration of how to use these released functions, we provide our liberally commented codes that were used to generate the numerical examples presented in Sec.~\ref{sec53} and \ref{sec55} of the present paper (see ``\textit{Examples 1}'' and ``\textit{Example 2}'', respectively). The first example implements the case in which a null training sample (NTS) of the stochastic noise $\bn$ is available. The second example implements the case without an NTS of the stochastic noise $\bn$. In this case, the NTS is estimated from the RV series under test.
We plan to improve this current release with new settings, a parallelization scheme, and practical examples of implementation for the most recent RV data analyses in the near future.
}
\newpage


\section{Main notations}
\label{ann1}

\begin{table}[h!] 
\centering
 \caption{{Table of the main notations used in the paper.}}
   \label{tab1} 
   \def\arraystretch{1.3}
\begin{tabular}{|c|c|} 
\hline
        $Z,z $  & $Z:$ Scalar random variable,\\
                & $z:$  one realization of $Z$ \\[0.1cm] 
\hline
        ${\bf{z}} $ & Vector $[z_1,\hdots,z_N]^\top$ \\[0.1cm] 
\hline \hline
        ${\cal H}_0, {\cal H}_1$ & Null and alternative hypotheses  \\[0.1cm] 
\hline
        $\bx$ &  Time series under test\\[0.1cm] 
\hline
        ${\bf{s}}$ & RV planet signal under ${\cal H}_1$ \\[0.1cm] 
\hline
        $\bn$ & Stochastic colored noise of covariance matrix $\boldsymbol{\Sigma}$\\[0.1cm] 
\hline
        $\bw$ & {White Gaussian noise of variance $\sigma_\bw^2$}\\[0.1cm] 
\hline
        $\bd$ & Nuisance signal\\[0.1cm] 
\hline
        $N$ & Number of data points \\[0.1cm] 
\hline \hline
         ${\cal{T}}_L$ &  {Null training sample} \\[0.1cm] 
\hline
        $L$ &  Number of time series available in the NTS \\[0.1cm] 
\hline
        ${\bf c}$& Ancillary time series that can be used to model $\bd$ \\[0.1cm] 
\hline \hline
        ${\cal{M}}_\bs(\boldsymbol{\theta}_\bs)$ & Model of ${\bf{s}}$ depending on parameters vector $\boldsymbol{\theta}_\bs$   \\[0.1cm] 
\hline
        ${\cal{M}}_\bn(\boldsymbol{\theta}_\bn)$&Model of $\bn$ depending on parameters vector $\boldsymbol{\theta}_\bn$ \\[0.1cm] 
\hline
        ${\cal{M}}_\bd(\boldsymbol{\theta}_\bd)$& Model of $\bd$ depending on parameters vector $\boldsymbol{\theta}_\bd$\\[0.1cm] 
\hline
        & Estimate of parameter vector $\boldsymbol{\theta}$, \\
$\widehat{\boldsymbol{\theta}}, \widehat{\bdelta}$      & and perturbation interval of this parameter \\[0.1cm] 
\hline
 & {Parameters' prior distribution for $\widehat{\btheta}$} \\
$\pi$ & with scale parameter  $\widehat{\bdelta}$   \\[0.1cm] 
\hline \hline
        $\nu_k$ &  Frequencies \\[0.1cm]  
\hline
        $ \Omega $ & Indices set of considered frequencies \\[0.1cm] 
\hline
        $\bf p$ &  Periodogram \\[0.1cm] 
\hline
        ${\bf{\overline{p}}}_L$ &  Periodogram averaged with $L$ training datasets \\[0.1cm] 
\hline
        ${\bf{\widetilde{p}}\;|\;{\bf{\overline{p}}}}_L$ &  Periodogram standardized by $\overline{\bf{p}}_L$ \\[0.1cm] 
\hline \hline
        $T_M$ & Test of the highest periodogram value (Max test) \\[0.1cm] 
\hline
        $T_C$ & Test of the $N_C^{\rm th}$ highest periodogram value \\[0.1cm] 
\hline \hline
        $(\textsf{P},\textsf{T})$ & Couple $\{$periodogram, test$\}$  \\
\hline
        $\Phi_{T}$ & CDF of the considered test statistic $T$ \\[0.1cm] 
\hline \hline
$t$ & Test statistic \\[0.1cm] 
\hline \hline
        $\{B,b\}$ & Number of Monte Carlo simulations in algorithm~\ref{fig_algo3} \\[0.1cm] 
\hline
$p_v$ & $p$-value  \\[0.1cm] 
\hline
\end{tabular}
\label{default}
\end{table}

\newpage
\section{Validation of MHD simulations with a ground-based RV dataset}
\label{appB}

The realism of the RVs extracted from 3D MHD simulations has been demonstrated in \citetads{2020A&A...635A.146S} by comparing them with spatial RV data taken from the GOLF/SoHO spectrophotometer (see Sec. 2.2.4 and Fig. 2 of \citeads{2020A&A...635A.146S}). In this appendix, we extend this demonstration to solar ground-based RV data taken with HARPS-N (\citeads{2019MNRAS.487.1082C}, \citeads{2021A&A...648A.103D}).{
We selected the $L=26$ synthetic MHD time series of solar granulation that are described in  Sec.~\ref{sec421} and the  $430$-day solar time series of HARPS-N .
Their estimated power spectral densities are shown in Fig.~\ref{fig0}.} 
The match (amplitudes and frequency dependence of the noise) of the PSDs in the period regime where granulation dominates (red horizontal arrow) is very good. This example shows that MHD simulations, tuned for the stellar type of a target host star, could be used to provide realistic time series of the stellar granulation noise of this star.

\begin{figure}[h!]
\resizebox{\hsize}{!}{\includegraphics{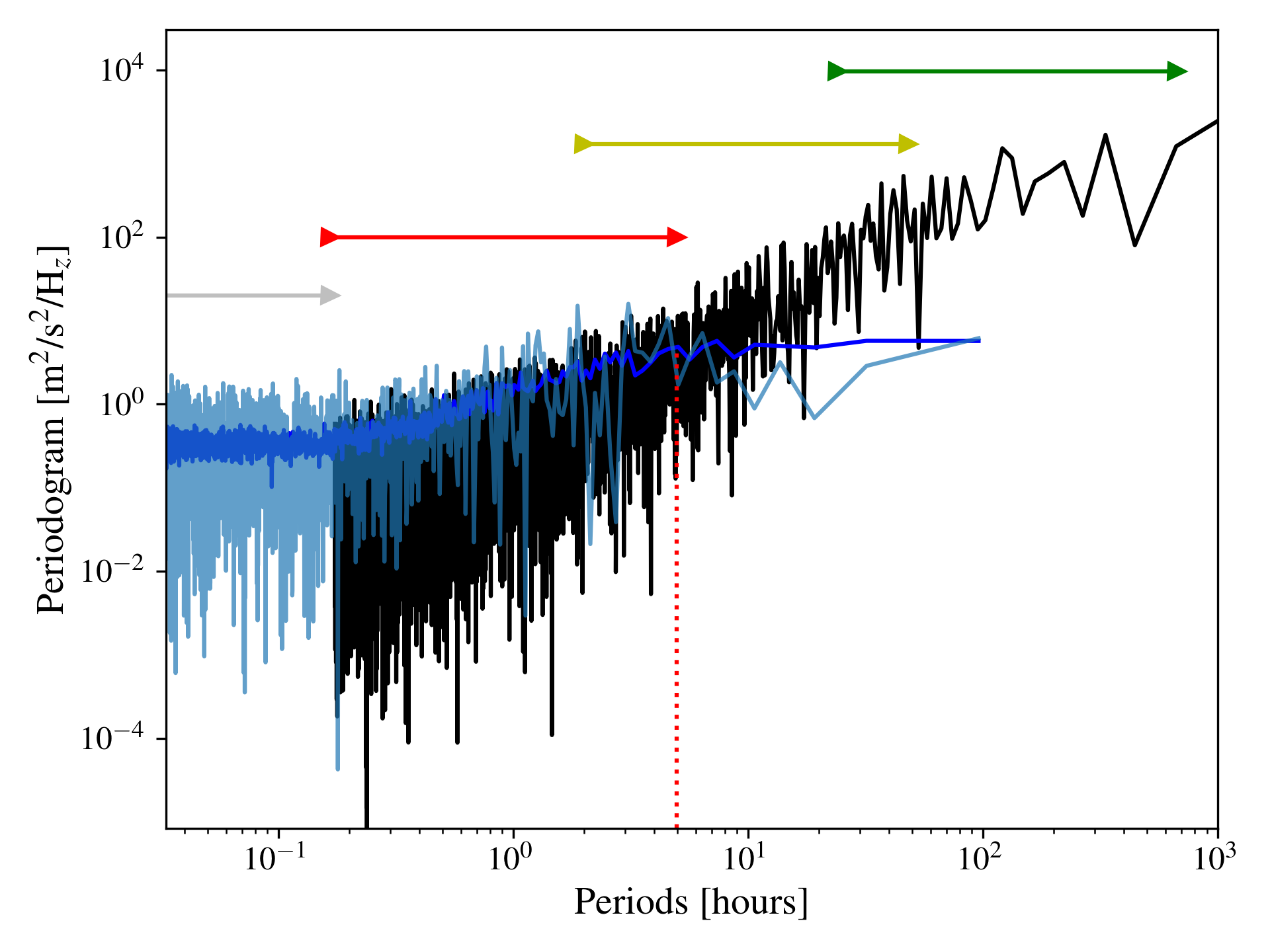}}
\caption{PSD comparison of solar RVs taken with the HARPS-N telescope (black, $430$ days) and synthetic RV extracted from 3D MHD simulations of solar granulation (light blue, $2$ days). The averaged periodogram of the $26$ MHD $\text{two}$-day time series is shown in dark blue. Period ranges dominates by high-frequency noise (instrumental and oscillations), granulation, supergranulation, and magnetic activity are indicated with gray, red, yellow, and green arrows.  The dotted red vertical line indicates the period limit ($\sim 5$ hours) within which the PSD is no longer dominated by the granulation noise.
}
\label{fig0}
\end{figure}

 \end{appendix}

\end{document}